\documentclass[onecolumn,superscriptaddress,nofootinbib, floatfix, amsfonts, aps]{revtex4}
\usepackage{amsmath}
\usepackage{bm}
\usepackage{graphicx}
\usepackage{mathrsfs}
\usepackage{footmisc}
\usepackage{multirow}
\usepackage{graphicx}
\usepackage[colorlinks, linkcolor=red,anchorcolor=green,citecolor=blue]{hyperref}
\usepackage{booktabs, ctable}
\usepackage{soul,xcolor}
\usepackage{xcolor, ulem, color, framed}
\usepackage{amssymb}

\begin{document}

\title{Anisotropic flows of charmonium in the relativistic heavy-ion collisions}

\author{Chenyu Li}
\affiliation{Culver Academies, Academy Rd, Indiana 46511, USA}

\author{Baoyi Chen}\email{baoyi.chen@tju.edu.cn}
\affiliation{Department of Physics, Tianjin University, Tianjin 300354, China}

%\date{\today}

\begin{abstract}
We review recent studies about anisotropic flows ($v_1, v_2, v_3$) of charmonium in the quark-gluon plasma produced in relativistic heavy-ion collisions. 
Collective flows of the bulk medium 
are developed due to the anisotropic pressure gradient 
of the medium. Strongly coupled with the bulk medium, charm quarks carry collective flows from the expanding medium, which will 
be inherited by the regenerated charmonium via the 
coalescence process. In event-by-event collisions where nucleon positions fluctuate from the smooth distribution, there is 
triangularity in the medium initial energy density. Triangular flows of the bulk medium and 
heavy flavor particles can be developed due 
to the initial fluctuations. In the longitudinal direction,  
the rapidity-odd distribution of the initial energy 
density is induced by the rotation of the medium in non-central heavy-ion collisions. 
Charmonium suffers biased dissociation along 
positive and negative $x$-directions in forward (backward) rapidity. The directed flow of charmonium becomes non-zero. 
The directed, elliptic and triangular flows ($v_1, v_2, v_3$) of charmonium 
come from the anisotropic initial distributions of the medium energy density in the transverse and longitudinal directions.

\end{abstract}
%\pacs{ }
\maketitle

\section{Introduction}
In the relativistic heavy-ion collisions,  
an extremely hot deconfined medium 
consisting of quarks and gluons, also called ``Quark-Gluon Plasma'' (QGP), 
is believed to be generated 
in the collision area~\cite{Bazavov:2011nk}. 
Studying the signals and the properties 
of the deconfined matter, which is like the state of our early 
universe is one of the important tasks in relativistic heavy-ion collisions. The 
deconfined matter is created when the 
energy density is above a critical value 
$\sim 1\ \rm{GeV/fm^3}$. Then the hot medium expands outside violently due to a large 
gradient of the energy density, and 
turns into a hadronic medium with a 
first-order or crossover phase transition 
when the 
temperature drops below a critical temperature $T_c$~\cite{Aoki:2006we}. 
The anisotropy of the energy density profile 
is transformed into the anisotropy of the azimuthal distribution of light hadrons~\cite{Heinz:2013th, Kolb:2003dz}. 
Based on the theoretical and experimental studies about the 
anisotropic flows of light hadrons~\cite{Song:2010mg,Gale:2012rq,STAR:2004jwm,ALICE:2011ab}, QGP turns out 
to be nearly a perfect fluid with a small ratio of shear viscosity over entropy density $\eta/s$. 
Heavy quark and quarkonium which are produced at the 
beginning of nuclear collisions, has been suggested to be a clean probe of an early stage of the medium~\cite{Matsui:1986dk}. Compared with  light hadrons which are usually produced 
at the hadronization hypersurface, heavy quarkonium is more sensitive 
to the initial profiles of the hot medium.  The color 
screening effect and random scatterings with thermal partons dissociate heavy quarkonium 
states in the hot medium. It suppresses the nuclear modification factor $R_{AA}$, which is defined to be the ratio of the quarkonium production in AA collisions and the production in 
proton-proton (pp) collisions scaled by the number of binary collisions $N_{coll}$~\cite{Braun-Munzinger:2000csl,Zhu:2004nw,Yan:2006ve,Liu:2010ej,Grandchamp:2003uw,Zhao:2011cv,Blaizot:2015hya,Yao:2020xzw,Krouppa:2015yoa,Yao:2020eqy,Wen:2022utn,Shi:2017qep}.  The suppression of $R_{AA}$ is proportional to the cold and 
hot nuclear matter effects, which have been observed in the experiments at 
the Super Proton Synchrotron
(SPS)~\cite{NA50:2000mfb}, the 
Relativistic Heavy-Ion Collider (RHIC)~\cite{PHENIX:2006gsi} and the Large Hadron Collider (LHC)~\cite{ALICE:2015jrl,ALICE:2016flj}. 

Charmonium nuclear modification factor is enhanced in AA collisions at the LHC energies where the medium temperatures are higher than the situation at RHIC energies. 
The increase of charmonium 
$R_{AA}$ at LHC energies is suggested to 
come from the coalescence of charm and anti-charm quarks 
in the QGP~\cite{Thews:2000rj,Greco:2003vf,Andronic:2003zv,Fries:2008hs,Du:2015wha,Chen:2017duy,Zhao:2017yan}. This coalescence contribution is proportional to the square of charm pair number $N_{c\bar c}$ in the medium, and is 
significantly enhanced when $N_{c\bar c}$ becomes large.  At LHC collision energies, 
$N_{c\bar c}$ is much larger than the value 
at RHIC and SPS energies. 
The $p_T$ spectrum of regenerated charmonium depends on the distribution of charm quarks 
in phase space. 
In the expanding QGP, charm quarks are strongly coupled with the medium to 
dump energy and carry collective flows 
in the medium~\cite{ALICE:2013olq,CMS:2012bms,Cao:2013ita,Cao:2018ews,He:2019vgs,He:2012df,Chen:2016mhl}. Heavy quark energy loss is 
determined by the coupling strength between heavy quarks and the medium. 
With stronger coupling strength, heavy quarks 
tend to reach momentum thermalization and diffuse into a larger volume with the 
expanding medium. Therefore, the regenerated charmonium from the kinetically thermalized charm quarks are mainly 
distributed at the low $p_T$ region~\cite{Liu:2009nb}. 
Besides, charm spatial diffusions with the expanding medium reduce the probability of their coalescence into 
a charmonium-bound state.  At the LHC 
energies where the regeneration mechanism dominates charmonium production, charmonium final distribution is closely connected with the dynamical evolutions of charm quarks in 
the medium.

In this manuscript, we review the recent studies about the effects of charm quark diffusions 
on the charmonium collective flows. To describe 
the dynamical evolutions of charm quarks 
and charmonium, the Boltzmann-type transport 
model~\cite{Yan:2006ve} and the Langevin-plus-coalescence 
model~\cite{Chen:2021akx} will be introduced. 
In the expanding medium, the diffusion of charm quarks reduces the spatial densities of charm 
quarks and suppresses the coalescence probability 
per one $c$ and $\bar c$. In semi-central 
collisions, the elliptic flows of the medium 
are developed. Additional triangular flows 
are also observed, which is induced by the 
fluctuations in the distribution of initial 
energy density.  The rotation of the hot medium 
also results in rapidity-odd distribution of 
the medium along the longitudinal direction. 
Charmonium suffers biased dissociation 
along the positive and negative $x-$directions 
in the tilted medium. The directed flow 
of charmonium is non-zero in forward (backward) 
rapidites. 
All of charmonium anisotropic flows in AA collisions are due to 
the anisotropic initial distribution of 
the medium energy density in 
transverse and longitudinal directions, and also 
the charm quark diffusions in the medium.

\section{charm thermalization and charmonium regeneration}
\subsection{Charm quark evolution}
As heavy quarks are stongly coupled with the quark-gluon plasma, they 
suffer significant energy loss and carry collective flows in the 
medium. Large collective flows of $D$-mesons have been observed in 
experiments where $D$-mesons are close to kinetic equilibrium when 
they move out of the medium. The dynamical evolutions of charm quarks 
in QGP and $D$-mesons in hadronic gas are regarded as Brownian motion. 
Heavy quarks lose energy via random scatterings with thermal partons 
and the gluon radiation induced by the medium. Including both 
hot medium effects, the time evolution of 
heavy quark momentum is described with the 
Langevin equation~\cite{Cao:2015hia}, 
 \begin{align}
\label{lan-gluon}
{d{\bf p}\over dt}= -\eta {\bf p} +{\bf \xi} + {\bf f_g}
\end{align}
whe ${\bf p}$ is the three-dimensional momentum of heavy quarks. 
The drag coefficient depends on the medium 
temperature $T$ and the momentum $\eta = \kappa/(2TE)$ with the 
definition of the energy $E=\sqrt{m_c^2+|{\bf p}|^2}$.
Charm quark mass is taken as $m_c=1.5$ GeV. 
The dynamical evolutions of the medium local 
temperatures $T({\bf x},t)$ will be 
given by the hydrodynamic model. 
The momentum diffusion coefficient $\kappa$ is 
connected with the spatial diffusion coefficient 
$\mathcal{D}_s$ via the relation $\kappa=2T^2/\mathcal{D}_s$. 
From theoretical and experimental studies about the elliptic flows and nuclear modification factors of $D$-mesons in nuclear collisions,  the value of $\mathcal{D}_s(2\pi T)$ is estimated to be $5\leq \mathcal{D}_s(2\pi T)\leq 7$ with weak dependence on temperature~\cite{Rapp:2018qla,Zhao:2020jqu}. It becomes larger in the hadronic medium due to a weaker coupling strength. 

In the noise term ${\bf \xi}$, 
the correlation between random scatterings at different time steps 
have been neglected, and ${\bf \xi}$ is approximated to be 
a white noise satisfying the relation, 
\begin{align}
\langle \xi^{i}(t)\xi^{j}(t^\prime)\rangle =\kappa \delta ^{ij}\delta(t-t^\prime)
\end{align}
where the momentum dependence have been neglected in ${\bf \xi}$. Three directions are represented by the index $i,j=(1,2,3)$. 
$t$ and $t^\prime$ are the time points of the evolution. 

In the third term, 
the gluon radiation induced by the medium is 
represented by the term $f_g=-d{\bf p}_g/dt$ which contributes 
a recoil force on heavy quarks. ${\bf p}_g$ is the 
three-dimensional momentum of an emitted gluon. In each 
time step $\Delta t$, the number of emitted gluons 
$\langle N_g\rangle$ in the time period $t\sim t+\Delta t$ 
is calculated with 
the formula~\cite{Cao:2015hia}, 
\begin{align}
\label{gluon-spec}
P_{\rm rad}(t,\Delta t) = \langle N_g(t, \Delta t)\rangle = \Delta t \int dx d k_T^2
{dN_g\over dx dk_T^2dt}
\end{align}
where the spectrum of emitted gluons are taken from 
Ref.\cite{Guo:2000nz,Zhang:2003wk}. 
When the time period $\Delta t$ 
is small enough, the number of emitted gluon 
$\langle N_g\rangle$ is always smaller than the unit, and the 
value of $P_{\rm rad}$ can be interpreted as the probability of 
emitting one gluon. 
$x=E_g/E$ is the ratio of emitted gluon and the heavy quark. 
$k_T$ is the transverse 
momentum of the emitted gluon. The radiation term dominates 
the energy loss of heavy quarks at high $p_T$  
region, while at low $p_T$ regions, heavy quark energy loss 
is dominated by the first two terms on the R.H.S of 
Eq.(\ref{lan-gluon}). In this work, we focus on the 
regeneration of charmonium via the coalescence of 
charm and anti-charm quarks, where most of regenerated charmonium 
are located at low $p_T$ bins.

The initial momentum distribution of heavy quarks 
can be calculated with the perturbative QCD model, such as PYTHIA and FONLL model~\cite{Cacciari:2001td}. The initial 
spatial distribution of heavy quarks is proportional to the 
product of two thickness functions, $dN_{c\bar c}/d{\bf x}_T=\sigma_{pp}^{c\bar c}
T_A({\bf x}_T-{\bf b}/2)T_B({\bf x}_T+{\bf b}/2)$. $\sigma_{pp}^{c\bar c}$ is the production cross-section of charm pairs in proton-proton collisions. $T_{A(B)}=\int dz \rho_{A(B)}({\bf x}_T, z)$ is the thickness function 
defined as the integration of the nucleon density over longitudinal direction. When 
heavy quarks move in the medium according to the Langevin equation, 
their positions are updated at each time step, 
${\bf x}(t+\Delta t)={\bf x}(t)+{\bf p}/E\cdot \Delta t$. 
Charm quark mass is larger than the medium temperature. The 
thermal production of charm pairs can be neglected and the total 
number of charm pairs are conserved when they do Brownian 
motion in the medium. Heavy quarks carry the information of 
the initial stage of the hot medium.

\subsection{Hot medium evolution}

\begin{figure}[!hbt]
\centering
\includegraphics[width=0.6\textwidth]{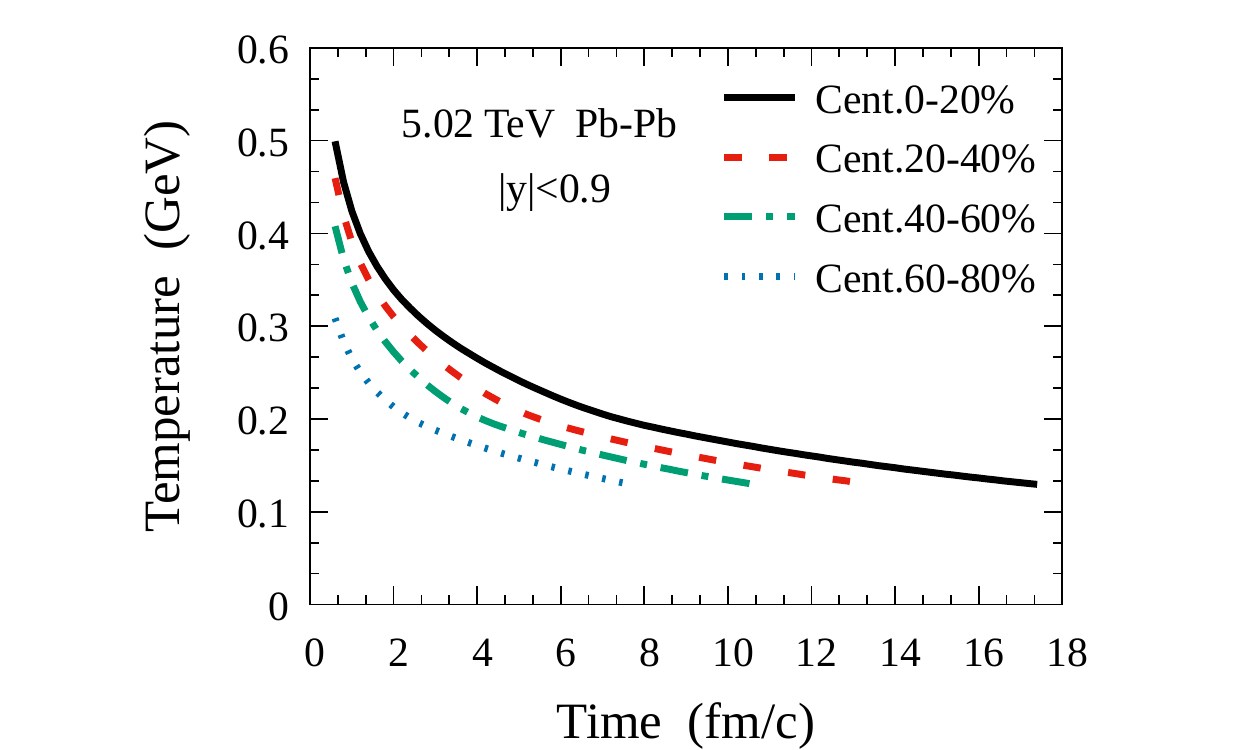}
\caption{
The time evolution of the local temperature at the 
center of the medium in the central rapidity in 
$\sqrt{s_{NN}}=5.02$ TeV Pb-Pb collisions. The collision centrality is chosen as 0-20\%, 20-40\%, 40-60\% 
and 60-80\%. This temperature evolution is calculated with the MUSIC model. This figure is cited from Ref.\cite{Chen:2021akx}. }
\label{lab-fig-temp}
\hspace{-0.1mm}
\end{figure}

To explain 
the large collective flows of light hadrons in experiments, 
the QGP medium is suggested to be a nearly 
perfect fluid. The dynamical evolution of the 
medium temperatures and the velocities are well described  
in the hydrodynamic models. The equation of state (EoS) 
of the QGP and hadronic medium are taken from 
lattice QCD calculation and the Hadron Resonance Gas model 
respectively~\cite{Huovinen:2009yb}.
In Fig.\ref{lab-fig-temp}, time evolution of the temperatures 
at the center of the medium are given by MUSIC model~\cite{Schenke:2010nt}. 
The collision energy is taken as  
$\sqrt{s_{NN}}=5.02$ TeV. The initial energy density profile 
is given by the Glauber model, where the maximum initial 
temperature is determined with the hadron multiplicity 
observed in 
heavy-ion collisions. The evolution of the temperature stops 
when it decreases to 
a kinetic freeze-out temperature $\sim 0.12$ GeV. 

\section{Charmonium thermal production}

\subsection{Diffusion of charm quarks}

With multiple charm quarks doing Brownian motion in the medium, uncorrelated $c$ and ${\bar c}$ which are 
usually from different $c\bar c$ pairs, may collide 
with each other to form a new charmonium state. At low temperatures 
close to $T_c$, heavy quark potential is 
partially restored, the newly formed charmonium state may survive from 
the hot medium. At LHC energies where the number of charm pairs is large, 
the coalescence of $c$ and $\bar c$ even dominates the final 
production of $J/\psi$ and $\psi(2s)$ in Pb-Pb collisions. 
The coalescence probability between one random $c$ and $\bar c$ quarks 
is written as

\begin{align}
\label{eq-psicoal}
&\langle \mathcal{P}_{c\bar c\rightarrow \psi}({\bf x_M}, {\bf p_M})\rangle_{\rm events} 
=g_M \int d{\bf x_1} d{\bf x_2}
{d{\bf p_1}\over (2\pi)^3} {d{\bf p_2}\over 
(2\pi)^3} {d^2N^{\rm norm}_1\over d{\bf x_1}d{\bf p_1}} 
{d^2N^{\rm norm}_2\over d{\bf x_2} d{\bf p_2}} 
W({\bf x_r}, {\bf q_r})   \nonumber \\
&\qquad\qquad \qquad \qquad\qquad\times
\delta^{(3)}({\bf p_M} -{\bf p_1}-{\bf p_2}) 
\delta^{(3)}({\bf x_M} - {{\bf x_1} +{\bf x_2}\over 2}), 
\end{align}
where ${\bf x}_{1,2}$ and ${\bf p}_{1,2}$ are the positions and 
momentum of $c$ and $\bar c$ quarks. 
${d^2N^{\rm norm}_1\over d{\bf x_1}d{\bf p_1}}$ is the normalized distribution of one 
charm in phase space. 
Delta function ensures the momentum conservation 
in the reaction $c+\bar c \rightarrow \psi +g$, 
where the gluon momentum has been neglected for simplicity. 
${\bf x}_M=({\bf x}_1+{\bf x}_2)/2$ and 
${\bf p}_M={\bf p}_1+{\bf p}_2$ are the position and momentum of the 
formed meson. The degeneracy factor 
$g_M=1/12$ is from the color and spin degrees of freedom. The 
Wigner function $W$ 
is the Weyl transform of the charmonium wave function, 
\begin{equation}
\label{wigner1}
W({\bf r},{\bf q})=\int d^3{\bf y} e^{-i{\bf q}\cdot{\bf y}}\psi\left({\bf r}
+{{\bf y}\over 2}\right)\psi^*\left({\bf r}-{{\bf y}\over 2}\right)
\end{equation}
where $\psi({\bf r})$ is charmonium wave function calculated from the Schrodinger equation with the in-medium 
heavy quark potential. When the wave function is approximated to 
be a Gaussian function, the corresponding Wigner function 
becomes $W({\bf x_r}, {\bf q_r})
= 8\exp[-{{x_r}^{2}\over \sigma^2} - \sigma^2 {q_r}^2]$~\cite{Greco:2003vf}. $x_r$ and 
$q_r$ are the relative coordinate and momentum between $c$ and 
$\bar c$ in the center of mass frame of charmonium. The information of the charmonium wave function is encoded in the  
Gaussian width, which is connected with the root-mean-square radius of charmonium,  $\sigma^2= {4\over 3}{(m_1+m_2)^2\over m_1^2 +m_2^2}
\langle r^2\rangle_M $~\cite{Greco:2003vf}.  The production $N_{M}^{AA}$ 
of charmonium in AA 
collisions is proportional to the square of charm pair number 
$N_{c\bar c}^{AA}$~\cite{Chen:2021akx}, 
\begin{align}
\label{eq-hadron}
&{ N_{M}^{AA}} = 
\int d{\bf x_M} { d{\bf p_M}\over (2\pi)^3}
\langle{\mathcal{P}_{c\bar c\rightarrow \psi}({\bf x_M},{\bf p_M})}\rangle_{\rm events}
%\nonumber\\&\qquad\qquad\qquad \times 
{( N_{c\bar c}^{AA})^2 } ,\\
& N_{c\bar c}^{AA} = \int d{\bf x_T} T_A({\bf x_T} -{{\bf  b}\over 2})
T_B({\bf x_T} +{{\bf  b}\over 2}) 
%&\qquad \qquad  \times
\mathcal{R}_S 
%\nonumber \\&\qquad\qquad\qquad \times 
{d\sigma_{pp}^{c\bar c}\over dy} \Delta y_{c\bar c},
\end{align}
where the factor $\mathcal{R}_S$ represents the nuclear shadowing effect. 
It changes the initial 
distribution of charm pairs in AA collisions, 
especially at the LHC energies. 

From Eq.(\ref{eq-psicoal}-\ref{eq-hadron}), charmonium final production depends on the dynamical evolutions of charm quarks in the medium. 
Strongly coupled with the medium, charm quark 
density is reduced by the expansion of the medium. 
This indicates that, in the hotter medium, it takes a longer time 
to cool down and restore the heavy quark potential. As a result, the charm quarks diffuse into a larger volume before 
they combine to form a charmonium state. 
This diffusion effect will suppress the coalescence probability of {\itshape one} $c$ and {\itshape one} $\bar c$.

%The ratio of the $J/\psi$ 
%nuclear modification factor $R_{AA}(5.02TeV)/R_{AA}(2.76TeV)$ 
%is larger than unit. It explains well the experimental data in %Fig.\ref{lab-raa-ratio}.

\begin{figure}[!hbt]
\centering
\includegraphics[width=0.45\textwidth]{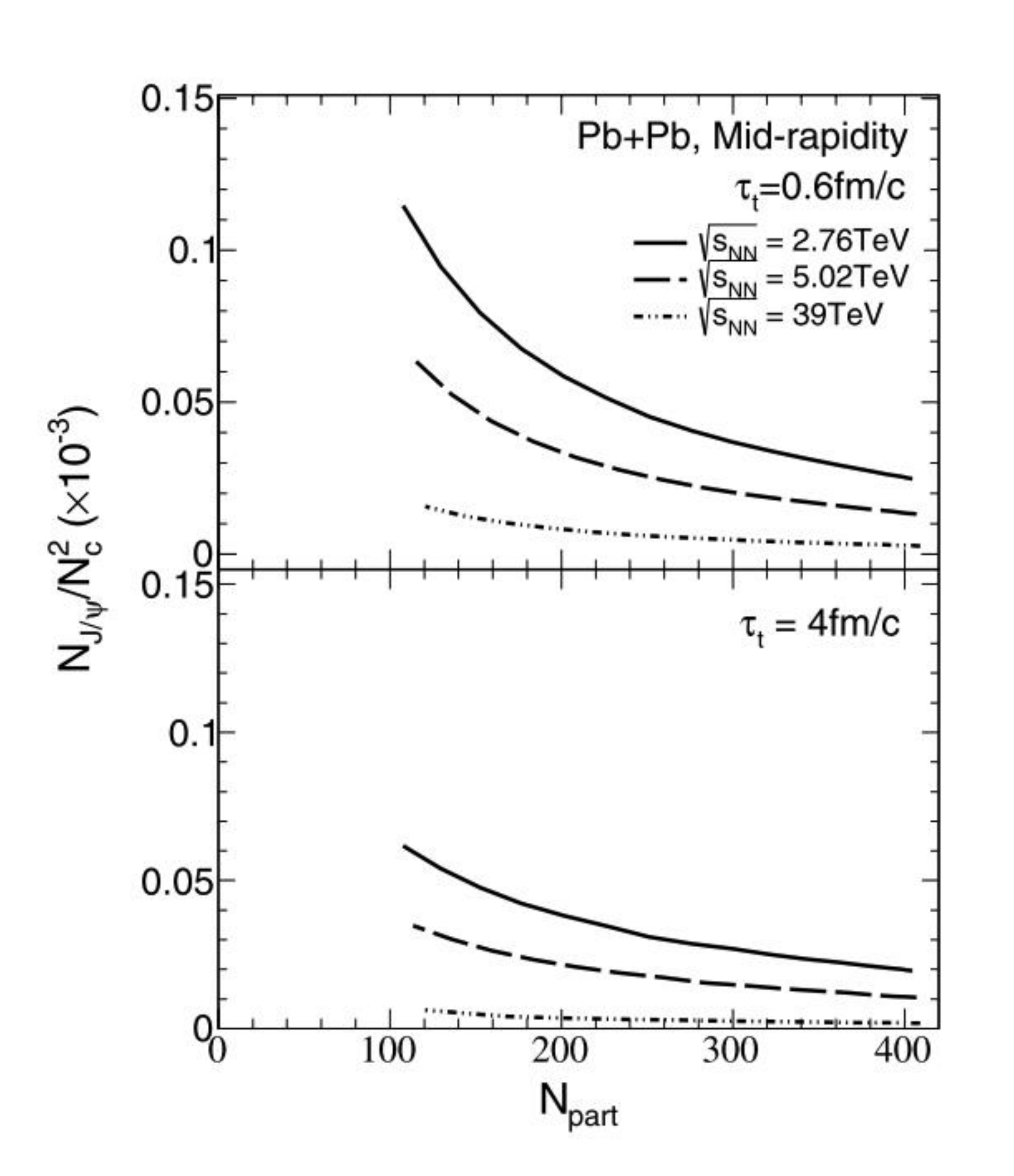}
\caption{
The ratio of $J/\psi$ production to the square of total charm pair number (with $N_c=N_D+N_{\Lambda_c}+...$ including all charm 
flavor hadrons) as a function of the number of participants $N_{part}$ in Pb-Pb 
collisions. The collision energies are chosen as $\sqrt{s_{NN}}=2.76,5.02,39$ TeV, respectively. 
Charm quark momentum distribution is assumed to reach equilibrium 
at the beginning of medium expansion $\tau_t=0.6$ fm/c in the upper panel and $\tau_t=4$ fm/c in the lower panel, respectively. 
This figure is cited from Ref.\cite{Zhao:2017yan}.}
\label{lab-psi-D}
\hspace{-0.1mm}
\end{figure}

 The thermal production of $J/\psi$ at $\sqrt{s_{NN}}=2.76$ TeV and 5.02 TeV Pb-Pb collisions are studied based on the 
coalescence model in Ref.\cite{Zhao:2017yan}.  
At 5.02 TeV, $J/\psi$ $R_{AA}$ is enhanced due to more significant number of charm pairs in the medium, compared with the situation at 2.76 TeV. 
However, the coalescence probability per  
one $c$ an $\bar c$ is reduced when the time of the 
medium expansion is longer. This effect is 
characterized by the ratio $N_{J/\psi}/(N_{c\bar c})^2$ 
of $J/\psi$ production over the square of charm pair 
number, please see Fig.\ref{lab-psi-D}. 
The ratio $N_{J/\psi}/(N_{c\bar c})^2$ is calculated as a function of $N_p$ at the collision energies of $\sqrt{s_{NN}}=(2.76, 
5.02, 39)$ TeV. Different 
time scales of charm momentum thermalization are considered in the 
figure. In upper and lower panels of 
the figure, charm quarks are assumed to reach momentum thermalization at $\tau\sim \tau_t$. 
Charm quarks do free streaming before the time scale 
$\tau_t$ and diffuse according to 
the diffusion equation in $\tau>\tau_t$.
In more central collisions, the initial 
energy density of the medium becomes larger. It takes 
longer time for the medium to decrease to $T_c$, 
where the coalescence of $c$ and $\bar c$ happens. The spatial density of charm quarks is diluted to reduce the coalescence probability of 
{\itshape one} $c$ and 
$\bar c$. 
Similarly, at higher collision 
energies with the production of higher energy density, 
the diffusion of charm quarks is more vital, and charm 
spatial density is more reduced.

\begin{figure}[!hbt]
\centering
\includegraphics[width=0.4\textwidth]{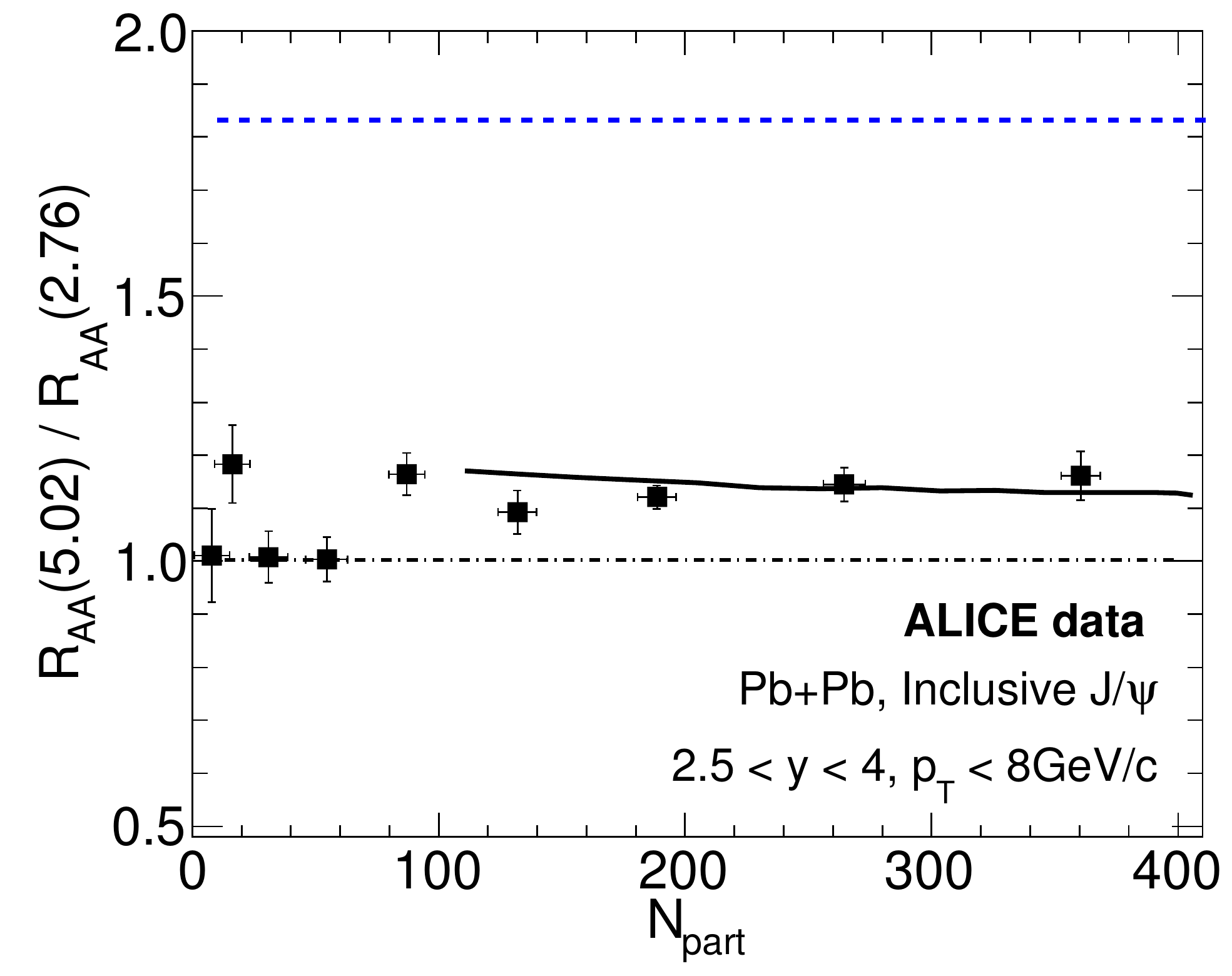}
\caption{
The ratio of inclusive nuclear modification factors $R_{AA}$ at 2.76 TeV and 5.02 TeV Pb-Pb collisions, 
plotted with a black solid line. 
Blue dotted line is a situation where 
same hydrodynamic evolution is used  as a test in both 
collision energies. The reduction from dotted to 
black solid lines is due to the stronger diffusion of 
charm quarks in the hotter medium produced at 5.02 TeV. 
This figure is cited from Ref.\cite{Zhao:2017yan}.}
\label{lab-raa-ratio}
\end{figure}

The charm diffusion in the expanding medium 
tends to suppress the nuclear modification factor $R_{AA}$. 
In Fig.\ref{lab-raa-ratio}, 
when employing the same hydrodynamic evolution 
at both $2.76$ TeV and $5.02$ TeV as a test, $R_{AA}(5.02 TeV)$ is enhanced because of a  
larger production cross section 
of charm pairs $d\sigma_{pp}^{c\bar c}/dy$ at 5.02 
TeV compared with the situation at 2.76 TeV. Their ratio 
is plotted with blue dotted line in Fig.\ref{lab-raa-ratio}. 
When taking the realistic evolutions for the medium produced 
at each collision energy, 
the coalescence probability 
per one $c$ and $\bar c$ is reduced at 5.02 TeV 
due to a stronger diffusion of charm quarks before 
the coalescence process. The 
difference between the blue dotted line (with the  
same hydrodynamic evolution) and the 
black solid line (with two realistic hydrodynamic 
evolutions) are due to the different diffusions 
of charm quarks in two collision energies.

\subsection{Elliptic flow $v_2$}

In non-central nuclear collisions, the collision area of two nuclei 
is elliptic. The gradient of the initial energy density 
becomes anisotropic in the transverse plane. With different 
acceleration in the $x$- and $y$-directions, the azimuthal 
distribution $dN/d\phi$ 
of the final hadrons is anisotropic as well, 
which can be 
expanded as, 

\begin{align}
    dN/d\phi \propto 1+2\sum_{n=1}^{\infty} v_n\cos n(\phi-\Phi_n)
    \label{eq-vn-def}
\end{align}
where $v_n$ and $\Phi_n$ are the magnitude and the phase 
of the $n$th-order harmonic flows in the amimuthal angle 
distribution of final particles. 
As charm quarks are strongly coupled with the medium, charm quarks 
diffuse in the expanding medium to carry elliptic flows. 
At the 
temperatures close to $T_c$ where heavy quark potential is 
partially restored, charmonium is regenerated via the coalescence 
of charm quarks and inherits the collective flows of  
charm quarks. 
 
In AA collisions, final charmonium comes from two 
mechanisms: 
the primordial production at the beginning of nuclear collisions 
and the regeneration from the coalescence of charm quarks 
in the QGP expansion. 
Charmonium bound states suffer dissociation 
in the hot medium. On the other hand, the regeneration process 
increases the charmonium final production. Both contributions 
can be included in the transport equation. Charmonium distribution 
in phase space $f_\Psi$ evolves with time. It can be 
simplified as~\cite{Zhu:2004nw,Yan:2006ve,Chen:2015iga,Chen:2013wmr}, 
\begin{equation}
\label{eq-trans}
\left[\cosh(y-\eta){\frac{\partial}{\partial\tau}}+{\frac{\sinh(y-\eta)}{\tau}}{\frac{\partial}{\partial
\eta}}+{\bf v}_T\cdot\nabla_T\right]f_\Psi
=-\alpha_\Psi f_\Psi+\beta_\Psi
\end{equation}
where $\tau$ and $\eta$ are the proper time and the pseudo-rapidity respectively. Here $y$ is the rapidity in momentum space. ${\bf v}_T={\bf p}_T/E_T={\bf p}_T/\sqrt{m_\Psi^2+p_T^2}$ 
is the transverse velocity of charmonium moving in the medium.
The terms on the L.H.S represents the charmonium distribution $f_\Psi$ changing with time $\tau$ 
and coordinates ($\eta,{\bf x}_T$). 
On the R.H.S, $\alpha$ and $\beta$ represent the 
decay rate and the regeneration rate of charmonium. 
The decay rate $\alpha_\Psi$ is proportional to the density of 
thermal gluons and the inelastic cross section. While 
the regeneration rate $\beta_\Psi$ 
in the reaction $c+\bar c\rightarrow \Psi+g$ is proportional to the densities of charm and 
anti-charm quarks and their coalescence probability. 
The formula of $\alpha_\Psi$ and $\beta_\Psi$ are given 
in Eq.(\ref{lab-alpha}-\ref{lab-beta}),
\begin{eqnarray}
\label{lab-alpha}
\alpha_\Psi({\bm p},{\bm x},\tau|{\bf b}) &=& {1\over 2E_T}\int{d^3{\bm p}_g \over (2\pi)^3 2E_g}W_{g\psi}^{c\bar c}(s)f_g({\bm p}_g,{\bm x},\tau)\Theta(T({\bm x},\tau|{\bm b})-T_c),\\
\label{lab-beta}
\beta_\Psi({\bm p},{\bm x},\tau|{\bm b}) &=& {1\over 2E_T}\int {d^3{\bm p}_g \over (2\pi)^3 2E_g}{d^3{\bm p}_c \over(2\pi)^3 2E_c}{d^3{\bm p}_{\bar c} \over(2\pi)^3 2E_{\bar c}}\nonumber\\
&&\times W_{c\bar c}^{g\psi}(s)f_c({\bm p}_c,{\bm x},\tau|{\bm b})f_{\bar c}({\bm p}_{\bar c},{\bm x},\tau|{\bm b})\nonumber\\
&&\times(2\pi)^4\delta^{(4)}(p+p_g-p_c-p_{\bar c})\Theta(T({\bm x},\tau|{\bm b})-T_c).
\end{eqnarray}

In the above formula, $f_g$ is the gluon density, 
$W_{g\psi}^{c\bar c}$ is the dissociation rate calculated from pQCD~\cite{Peskin:1979va, Bhanot:1979vb}.  
The step function $\Theta(T-T_c)$ ensures that the 
parton dissociation only exists in the QGP. 
In $\beta_\Psi$, $W_{c\bar c}^{g\psi}$ is 
connected with $W_{g\psi}^{c\bar c}$ via the 
detailed balance. $\delta$ functions 
indicate the energy-momentum conservation in the 
coalescence process. 
$f_c$ and $f_{\bar c}$ are the 
density of $c$ and $\bar c$ quarks. Their distribution 
can be obtained from the Langevin model. As $D$ meson 
elliptic flows 
are large and close to the flows of light hadrons~\cite{ALICE:2013olq,ALICE:2012vgf}, 
in the thermalization limit, 
charm spatial density satisfies the equation 
$\partial_\mu(\rho_c u^\mu)=0$, 
where $u^\mu$ is the four-velocity of the fluid. 
While 
charm momentum distribution is taken as a normalized 
Fermi distribution as an approximation.

 The initial distribution of primordially produced 
charmonium in nuclear collisions is obtained from the 
production in proton-proton collisions scaled with the 
number of binary collisions $N_{coll}$. In proton-proton 
collisions, the $p_T$-spectrum of $J/\psi$ can be 
parametrized with the form, 
\begin{align}
    \label{eq-pp1S}
&f_{pp}^{J/\psi}\equiv {d^2\sigma_{\mathrm{pp}}^{J/\psi}\over dy2\pi p_Tdp_T} = 
{d\sigma_{\mathrm{pp}}^{J/\psi}\over dy} 
\times {(n-1)\over {\pi(n-2)\langle p_T^2\rangle_{pp}}}
[1+{p_T^2\over {(n-2)\langle p_T^2\rangle_{pp}}}]^{-n},
\end{align}
where the rapidity-differential cross section $d\sigma_{pp}^{J/\psi}/dy$ 
is determined by experiments~\cite{ALICE:2012vup,ALICE:2012vpz}.  The parameter 
$n$ and $\langle p_T^2\rangle$ characterize the shape of 
the $J/\psi$ $p_T$-distribution in pp collisions~\cite{Chen:2018kfo}. In AA collisions, cold nuclear matter (CNM) effect 
also changes the production of charmonium. Nuclear shadowing 
effect modifies the parton densities in the nucleus which 
affect the charmonium production. Besides, the partons from 
nucleons scatter with other nucleons before fusing into 
a charmonium and charm pair. This process increases the $p_T$ of produced 
charmonium compared with the situation in pp collisions. It is also called 
Cronin effect, which can be included by modifying 
 $f_{pp}^{J/\psi}$ with a Gaussian smearing method. 
All of the cold nuclear matter effects happen before the 
hot medium effects. Therefore, those CNM effects are included 
in the initial distribution of Eq.(\ref{eq-trans})~\cite{Chen:2016dke},
\begin{align}
    f_\Psi({\bf x}, {\bf p}_T,y, \tau_0)=
    (2\pi)^3\delta(z)T_A({\bf x}_T-{\bf b}/2) 
    T_B({\bf x}_T+{\bf b}/2)\mathcal{R}_S^A\mathcal{R}_S^B
    \overline{f_{pp}^{J/\psi}}
\end{align}
Where the shadowing factor $\mathcal{R}_S^{A(B)}$ is calculated 
with the EPS09 NLO model~\cite{Eskola:2009uj}. $\overline{f_{pp}^{J/\psi}}$ is the 
momentum distribution of primordially produced $J/\psi$ with 
the modification of Cronin effect. 

\begin{figure}[!hbt]
\centering
\includegraphics[width=0.45\textwidth]{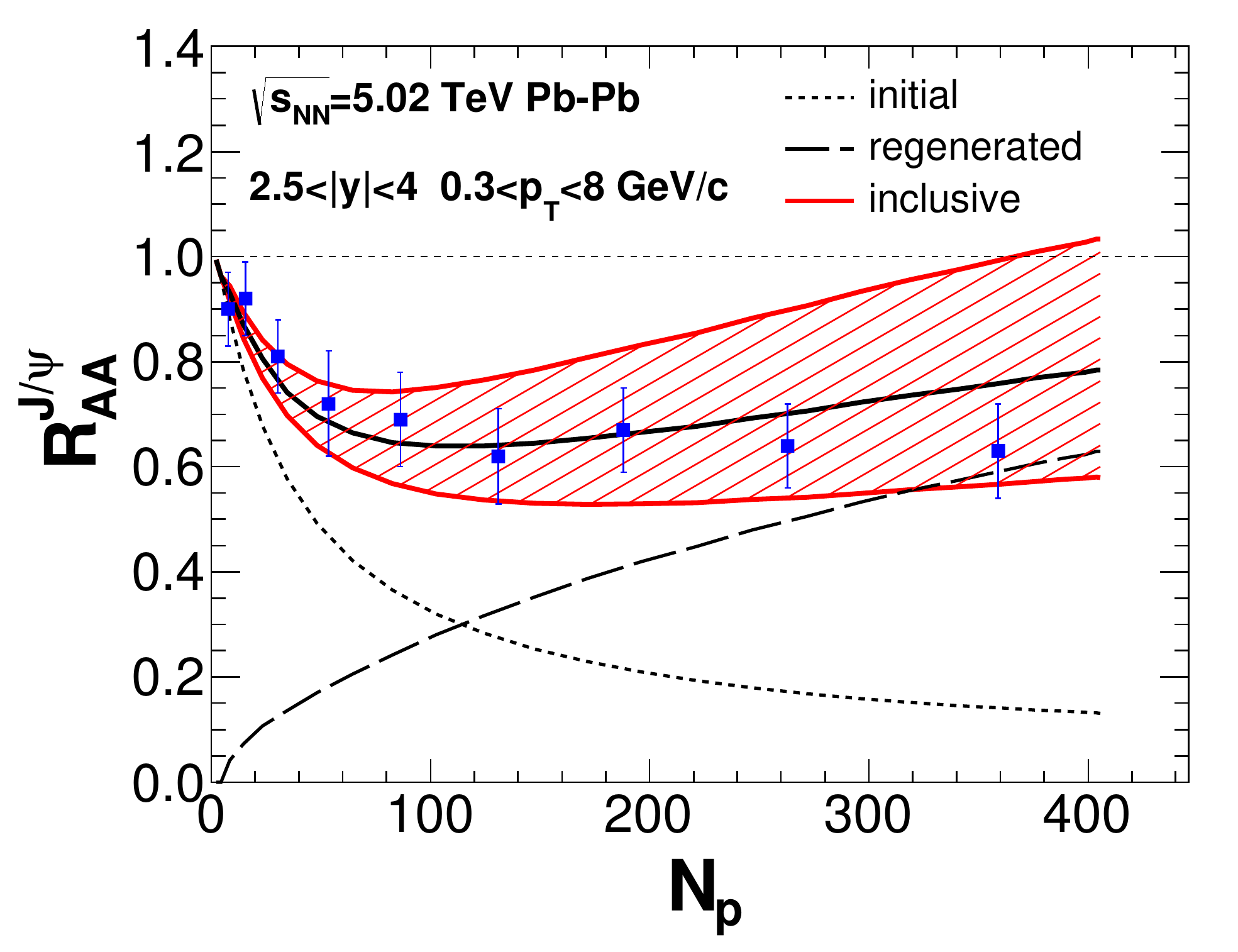}
\includegraphics[width=0.48\textwidth]{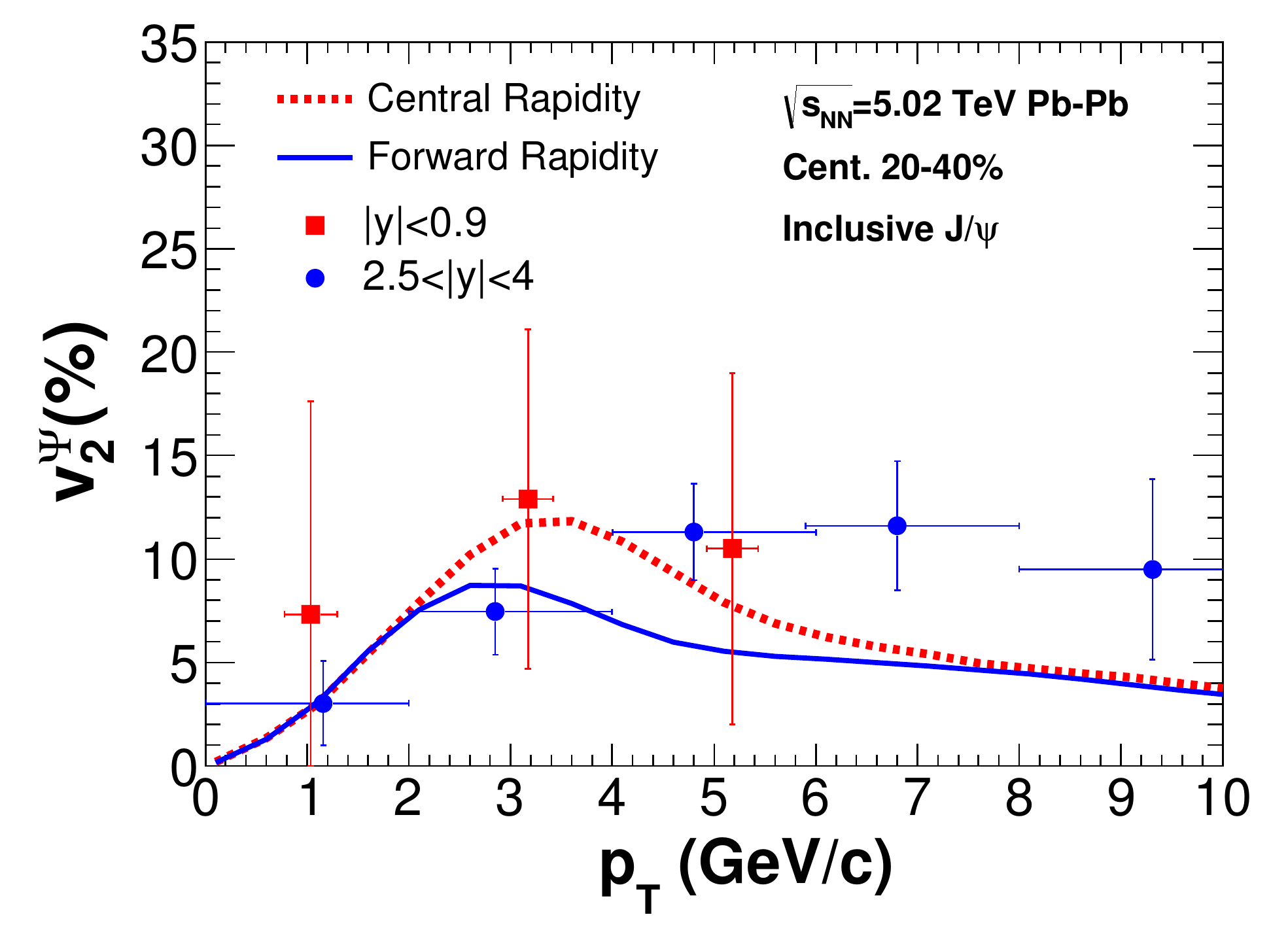}
\caption{
(Left panel) $J/\psi$ nuclear modification factor $R_{AA}$ as 
a function of number of participant $N_p$ at $\sqrt{s_{NN}}=5.02$ TeV Pb-Pb collisions. Dotted, dashed and solid lines represent 
the primordial production, regeneration and inclusive production, 
respectively. The band corresponds to the uncertainty in the 
production cross section of charm pair $d\sigma_{pp}^{c\bar c}/dy$. 
(Right panel) Elliptic flows of inclusive $J/\psi$ as a function of $p_T$ in centrality 0-20\%. 
This figure is cited from Ref.\cite{Chen:2018kfo}.}
\label{lab-v2-jpsi}
\hspace{-0.1mm}
\end{figure}

Solving the transport model, 
$J/\psi$ nuclear modification factor including both primordial 
production and regeneration is plotted in the left panel of  Fig.\ref{lab-v2-jpsi}. Dotted, dashed and solid lines 
represent the primordial production, regeneration, and inclusive 
production. In semi-central and central collisions, the 
regeneration dominates the final production of $J/\psi$. The regenerated 
charmonium are mainly located at low $p_T$ region because of the charm quark energy loss. The primordial production dominates the production at high $p_T$. 
Therefore, in the right panel of Fig.\ref{lab-v2-jpsi}, 
$v_2$ of final inclusive $J/\psi$ is significant at low $p_T$ regions 
because of the collective flows of charm quarks. 
At high $p_T$, $v_2$ is small. Because $J/\psi$ as a color-singlet state is not strongly coupled with the medium like single charm quarks. Primordially produced charmonium does free streaming in the medium. And the small non-zero $v_2$ of $J/\psi$ is induced by the path-length-difference of trajectories when they move 
along different directions in the transverse plane. At high $p_T$ region 
in the right panel of Fig.\ref{lab-v2-jpsi}, theoretical calculation is 
below the experimental data. This may be due to the non-thermal 
distribution of 
charm quarks in the medium~\cite{He:2021zej}.

\subsection{Triangular flow $v_3$}
In the event-by-event heavy-ion collisions, 
the nucleon positions fluctuate and deviate from the smooth distribution 
of the nucleus~\cite{Alver:2010gr,Miller:2003kd}. 
The energy density produced by the collisions also 
fluctuate event-by-event. The realistic fluctuating profiles of the 
medium energy density is simulated with the Monte Carlo Glauber model 
(MC-Glauber)~\cite{Miller:2007ri} and Monte Carlo fKLN model (MC-KLN)~\cite{Kharzeev:2001gp}. In the dynamical expansion 
where the anisotropy in the initial distribution of 
medium energy density is transformed into the 
anisotropy in the momentum distribution of hadrons, 
there is non-zero triangular flow in the azimuthal distribution of hadrons. 
In experiments, the $v_3$ of light hadrons have been 
observed and explained well with the hydrodynamic 
model plus the fluctuating initial conditions. 
The value of $v_3$ is proportional to the triangularity 
in the initial fluctuating distribution. 
The regenerated charmonium from the coalescence of 
charm quarks also carry collective flows. 
The magnitude of the charmonium $v_3$ is determined 
by the initial fluctuations and the degree of charm 
kinetic thermalization.

\begin{figure}[!hbt]
\centering
\includegraphics[width=0.6\textwidth]{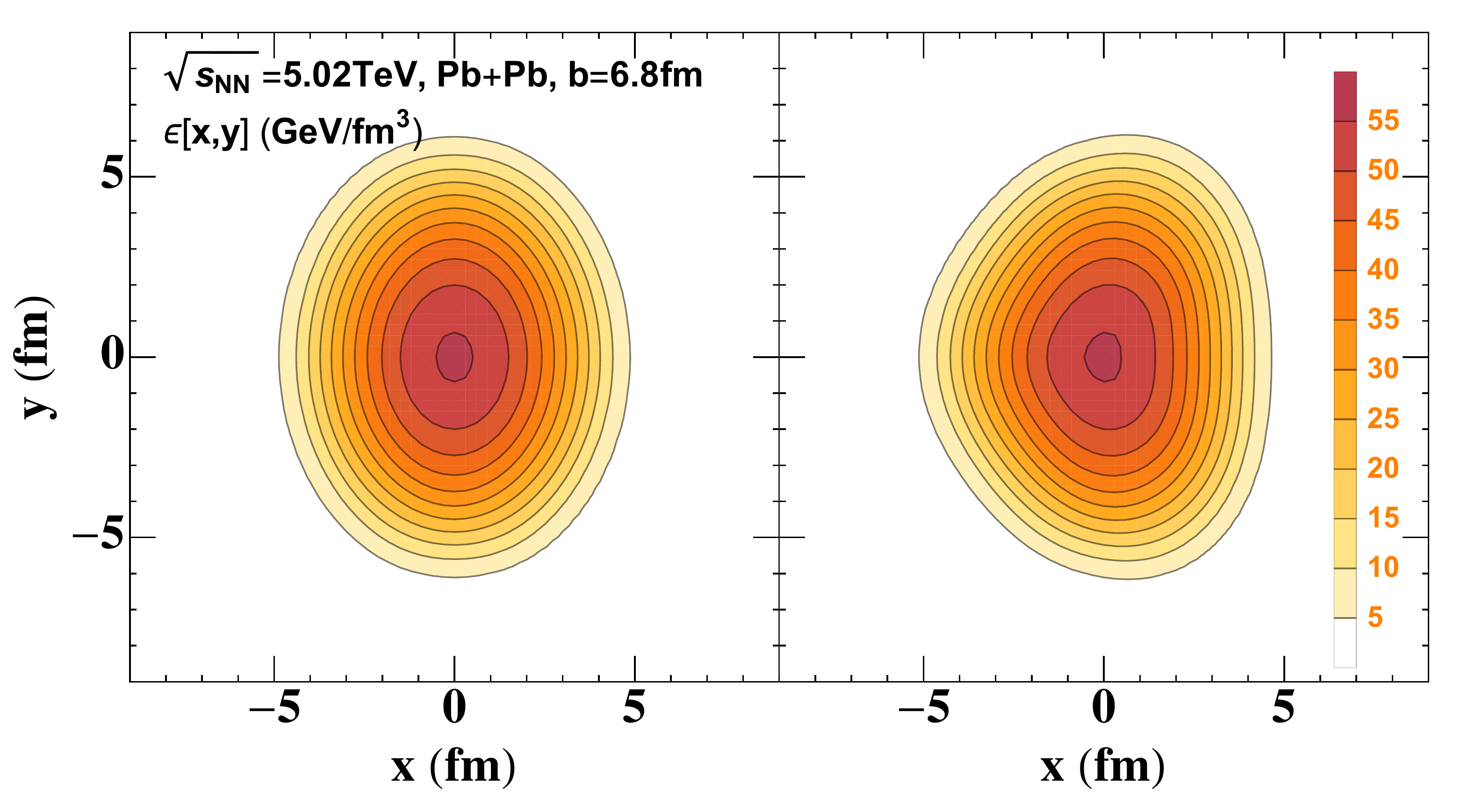}
\caption{
The initial energy density on the transverse plane in $\sqrt{s_{NN}}=5.02$ TeV Pb-Pb collisions. In the left panel, 
No fluctuations are considered. In the right panel, a 
deformation characterizing the triangularity in the 
energy density is included. 
This figure is cited from Ref.\cite{Zhao:2021voa}.}
\label{lab-v3-inite}
\hspace{-0.1mm}
\end{figure}

Firstly, one can generate a large set of fluctuating initial conditions with the Monte Carlo method. Then take 
these fluctuating medium evolutions into the transport model to calculate the $v_3$ of $J/\psi$ in each event. 
This calculation needs a huge computation time. 
Instead, 
one can take a single-shot hydrodynamics where triangular 
distributions of the initial energy density from fluctuations 
are encoded in the initial conditions. 
To introduce the triangularity in the initial energy 
density of the medium, a deformation factor is included 
in the formula of the energy denstiy~\cite{Alver:2010dn}, 
\begin{eqnarray}
\label{epsilon0+}
\epsilon({\bm x},\tau_0|{\bm b})\to \epsilon (\tilde{\bm x},\tau_0|{\bm b})
\end{eqnarray}
where the deformed coordinate is 
$\tilde{\bm x}=(x_T\sqrt{1+\varepsilon_3\cos[3(\phi_s-\Psi_3)]},\phi_s,\eta)$. 
$x_T=\sqrt{x^2+y^2}$ and $\phi_s=\arctan(|y|/|x|)$ are the transverse radius and azimuth angle.
$\Psi_3=0$ is regarded as the reaction plane angle in the 
smooth hydrodynamic evolution. 
The value of the parameter $\epsilon_3$ 
characterizing the magnitude of the fluctuation is determined 
to be $\epsilon_3=0.08,0.1,0.2$ in the centrality bin 0-10\%, 10-30\% and 30-50\% respectively 
in 5.02 TeV Pb-Pb 
collisions according to the simulation from MC-Glauber or MC-KLN model~\cite{Alver:2010dn,Qiu:2011iv}. 
The spatial 
distributions of the energy density with and without the fluctuations 
are plotted in Fig.\ref{lab-v3-inite}.

\begin{figure}[!hbt]
\centering
\includegraphics[width=0.3\textwidth]{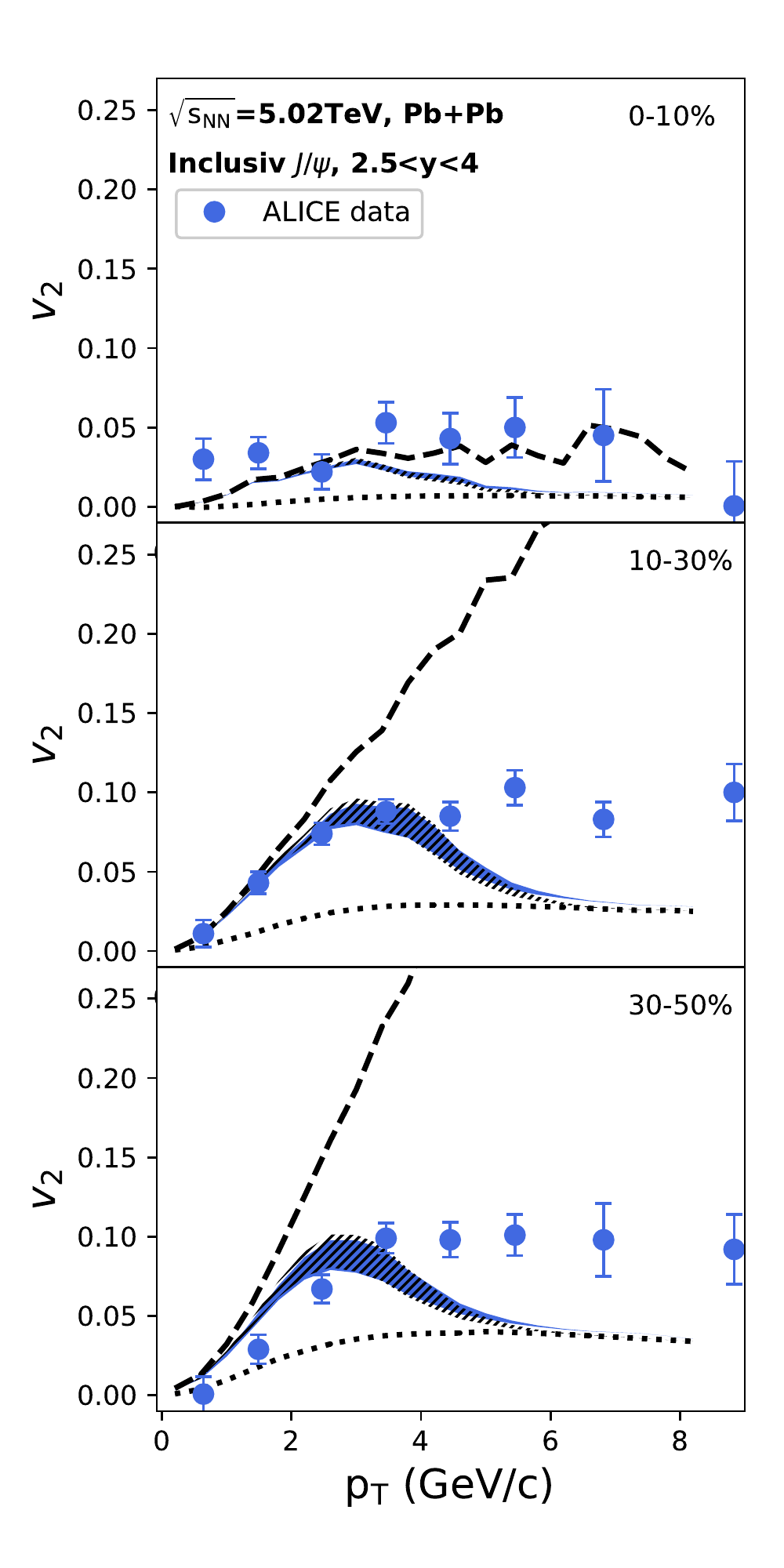}
\hspace{0.9cm}
\includegraphics[width=0.3\textwidth]{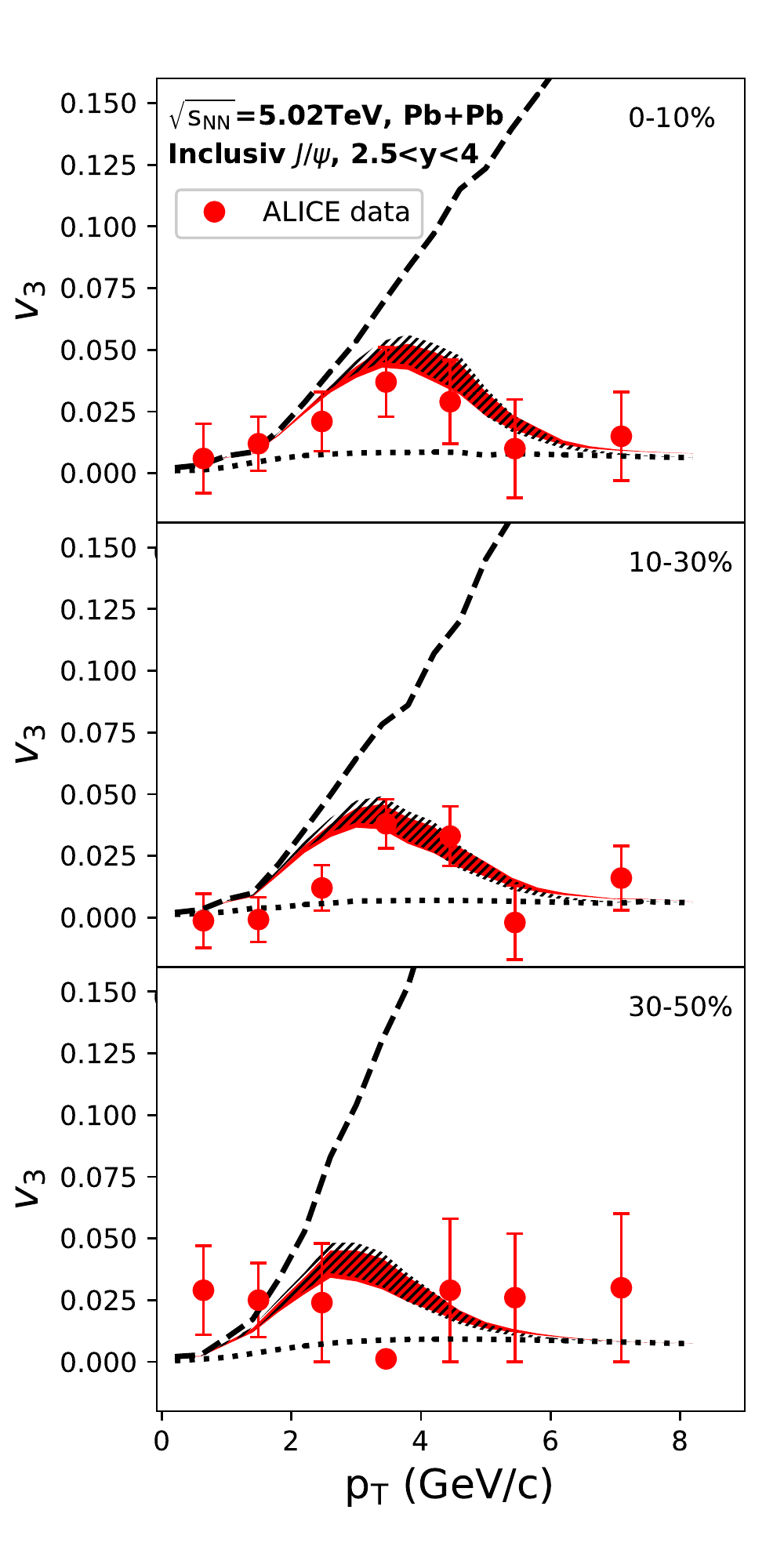}
\caption{ (Left panel) Elliptic flows of $J/\psi$ as a function 
of $p_T$ in $\sqrt{s_{NN}}=5.02$ TeV Pb-Pb collisions. 
The dotted, dashed lines and shadow bands 
represent the elliptic flows of 
initially produced, regenerated and inclusive $J/\psi$ respectively. 
The collision centrality is 0-10\%, 10-30\%, 30-50\% respectively. (Right panel) triangular flows of charmonium as 
a function of $p_T$. 
This figure is cited from Ref.\cite{Zhao:2021voa}.}
\label{lab-v3-pt}
\hspace{-0.1mm}
\end{figure}

In the left panel of Fig.\ref{lab-v3-inite}, 
the shape of initial energy density is elliptic which 
comes from the geometry size of the collision area. In the 
right panel of Fig.\ref{lab-v3-inite}, the 
triangular shape in the profile of the energy density 
is 
encoded which characterizes the magnitude of the 
fluctuations. 
The corresponding elliptic and triangular flows ($v_2,v_3$) 
of $J/\psi$ 
are calculated based on the transport equation 
by taking the 
corresponding hydrodynamic evolutions, please see 
Fig.\ref{lab-v3-pt}.

In Fig.\ref{lab-v3-pt}, elliptic flows depend sensitively 
on the collision centrality. In central collision, the 
pressure gradient of the medium energy density is nearly 
isotropic in the transverse plane. Elliptic flows of 
the medium and heavy flavor particles are  
small in cent.0-10\%. 
In cent.10-30\%, the elliptic flow become larger 
due to the elliptic profile of the collision area where 
medium acceleration is different along $x-$ and 
$y-$directions in the transverse plane. At high $p_T$ region, 
theoretical calculations underestimate the experimental data. 
This may be due to the non-thermal distribution of charm 
quarks~\cite{He:2021zej}. 
For triangular flows, they mainly come from the 
fluctuation of the medium 
energy density, which shows relatively 
weak dependence 
on the impact parameter. Charmonium $v_3$ in three 
centralities are close to each other 
without significant 
difference. 
The observation of $J/\psi$ $v_3$ indicates clearly that 
charm quarks are strongly coupled with the medium to 
be affected by the 
fluctuations.

\subsection{Directed flow $v_1$}

\begin{figure}[!hbt]
\centering
\includegraphics[width=0.32\textwidth]{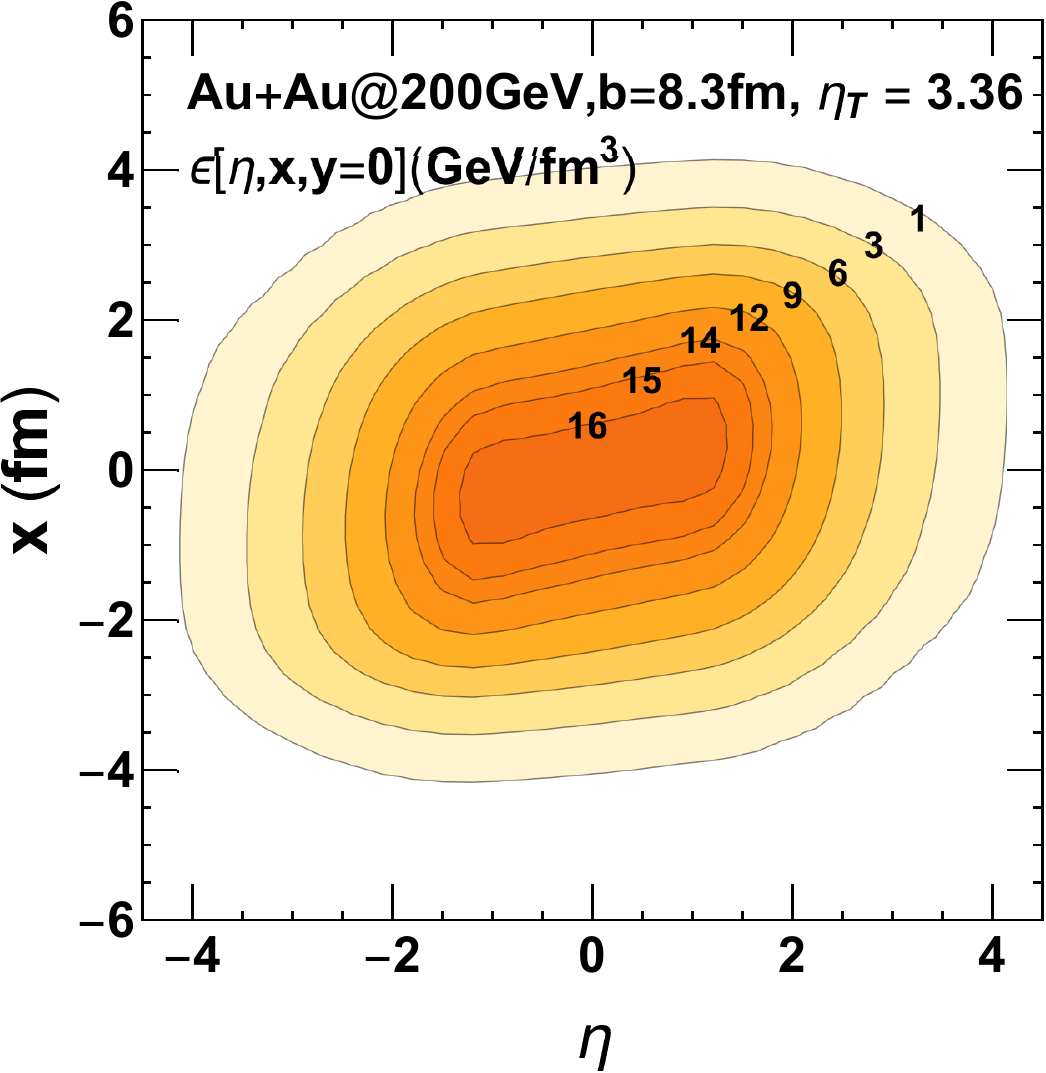}
\hspace{0.9cm}
\includegraphics[width=0.40\textwidth]{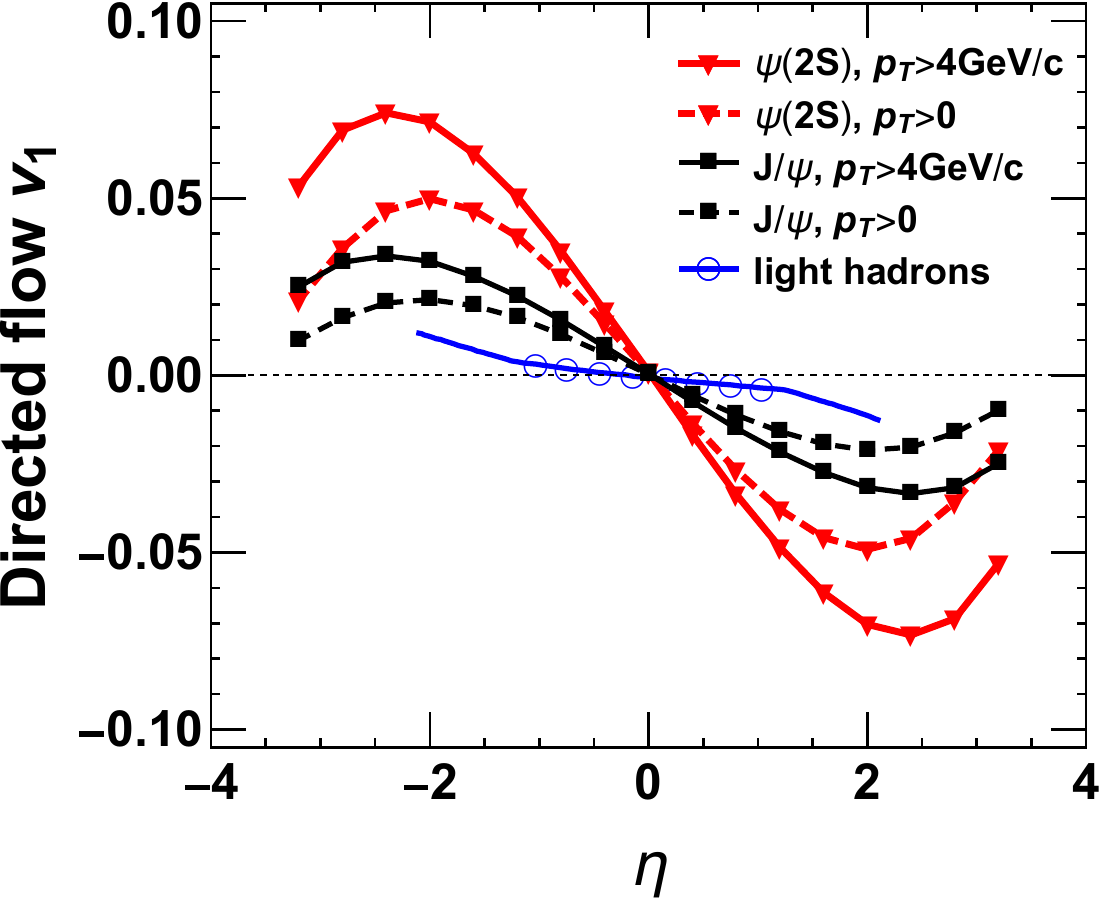}
\caption{ 
(Left panel) rapidity-odd distribution of initial 
entropy density in $\sqrt{s_{NN}}=200$ GeV Au-Au collisions. 
The collision centrality is 0-80\% corresponding to the 
averaged impact parameter $b=8.3$ fm. 
(Right panel) The directed flows of $J/\psi$ and $\psi(2s)$ 
as a function of rapidity $\eta$. Different $p_T$ bins ($p_T>0$ 
and $p_T>4$ GeV/c) are calculated. 
This figure is cited from Ref.\cite{Chen:2019qzx}.}
\label{lab-v1-all}
\hspace{0.1mm}
\end{figure}

In the previous text, we have discussed the anisotropy of the 
medium initial energy density in the transverse plane. 
In this section, we discuss the rapidity-odd distribution 
of the energy density induced by the rotation of the medium 
in semi-central nuclear collisions. 
In the collision area, 
The nucleus moving with the positive (negative) rapidity 
tends to tilt the produced bulk medium to positive (negative) 
$x$ with respect to the beam axis for positive (negative) 
$z$. The magnitude of the tilt in the bulk medium is connected 
with the directed flows of light-charged hadrons, which have 
been observed in experiments~\cite{STAR:2008jgm,STAR:2014clz}. In the Au-Au collisions at RHIC 
energy, the rapidity distribution of the medium energy 
density is plotted in the centrality 0-80\%, please see 
the left panel of Fig.\ref{lab-v1-all}. In the 
forward (backward) rapidity, 
the medium is tilted towards the positive (negative) 
$x$-direction. The initial entropy density of the medium 
is parametrized with the formula, 
\begin{align}
    s(\tau_0, {\bf x}_T, \eta) = s_0
&\times
\exp[-\theta(|\eta|-\eta_0) 
{(|\eta|-\eta_0)^2 \over 2\sigma^2}] \nonumber \\
&\times [ c_{hard} N_{coll} + (1-c_{hard}) 
(N_{part}^{+} \zeta_{+}(\eta)  
+ N_{part}^{-} \zeta_{-}(\eta))] 
\end{align}
where $s_0$ and the parameters in the 
exponential factor characterize the initial 
entropy density and the rapidity-odd distribution of 
the medium respectively~\cite{Chen:2019qzx}. The initial entropy 
density depends on both 
soft and hard process where the fractions are $(1-c_{\rm hard})$ and $c_{\rm hard}=0.05$ respectively. 
$N_{part}^+$ and $N_{part}^-$ 
are the number of participants in the forward and backward rapidities. 
The feature of the rapidity-odd distribution is
introduced via the function $\zeta_{+,-}(\eta)$. The 
hydrodynamic evolutions of the medium 
in both transverse and longitudinal 
directions are given by the MUSIC model with the input of the rapidity-odd 
initial entropy density.

The directed flows of $D$ mesons 
have 
been observed~\cite{STAR:2019clv}. 
Compared with light hadrons, charmonium are 
produced in the initial parton hard scatterings where the 
tilted shape of the entropy distribution is apparent. Besides, 
different from charm quarks, 
primordially produced charmonium with zero net color charge is weakly coupled with 
the medium. Therefore, 
charmonium suffer biased dissociation along 
positive and negative $x$-directions in the tilted medium, and 
this information is not contaminated by the following QGP 
evolutions where the tilted shape of the medium 
is less evident. In the right 
panel of Fig.\ref{lab-v1-all}, directed flows of $J/\psi$ and 
$\psi(2s)$ as a function of pseudorapidity 
$\eta$ are calculated 
in different $p_T$ bins in $\sqrt{s_{NN}}=200$ GeV 
Au-Au collisions~\cite{Chen:2019qzx}. As one can see, in the forward rapidity, 
charmonium moving along positive $x$-direction suffer 
stronger medium dissociation due to the longer path-length 
in the medium. 
This results in anisotropy in the azimuthal distribution 
of the final charmonium, where charmonium 
$v_1$ is negative in forward rapidity and negative in backward 
rapidity. 
For excited state $\psi(2s)$, the binding energy is smaller 
than the value of the ground state. 
They are more easily dissociated 
and sensitive to the profile of the medium 
entropy density. 
In Fig.\ref{lab-v1-all}, charmonium directed flows are bigger 
than the values of the light hadrons.  
In above calculations about $v_1$, the regeneration process has been neglected for simplicity, which becomes 
less important, especially at high $p_T$ region 
where the directed flow is strong, 
shown as the lines with $p_T>4$ GeV/c in Fig.\ref{lab-v1-all}.
Confirming and quantifying the charmonium 
$v_1$ help to extract the rotation and the 
rapidity-odd distribution of the hot medium.

\section{summary}
In the work, we review recent studies about charmonium 
anisotropic flows in relativistic heavy-ion 
collisions. The pressure gradients of the medium 
energy density become different along different directions, 
where the acceleration of the medium expansion is also anisotropic. 
With strong coupling between heavy quarks and the bulk medium, 
charm quarks approaches kinetic thermalization to carry 
collective flows from the expanding medium. 
The anisotropic flows of charm quarks will be inherited by 
the regenerated charmonium via the coalescence process. 
The elliptic flow of charmonium originates from  
the shape of the QGP produced in the collision 
area of two nuclei 
in semi-central collisions. It depends on the 
impact parameter and becomes almost zero in most 
central collisions. The 
triangular flow of charmonium comes from the 
fluctuations of the medium 
initial energy density. As 
the fluctuations of nucleon positions in the nucleus show weak dependence on the impact parameter, 
charmonium $v_3$ are similar to each other without 
evident difference in central and semi-central collisions. 
In the longitudinal direction, the rapidity-odd distribution 
of the 
energy density is introduced due to the rotation of the 
hot medium produced in non-central collisions. This results in 
biased charmonium dissociation along positive and negative $x$-directions due to the different path-length in QGP. The 
directed flows of $J/\psi$ and $\psi(2s)$ are larger than the 
case of light hadrons. As primordially produced charmonium 
is mostly dissociated by the medium in the initial stage where 
the tilted shape of the medium is evident, charmonium directed flow is proposed as a sensitive probe of the profile of the medium initial energy densitiy in relativistic heavy-ion collisions.

\vspace{0.5cm}
{\bf Acknowledge:}
This work is supported by the
National Natural Science Foundation of China (NSFC)
under Grant No. 12175165.


\begin{thebibliography}{20}





\bibitem{Bazavov:2011nk}
A.~Bazavov, T.~Bhattacharya, 
\textit{et al.}
%``The chiral and deconfinement aspects of the QCD transition,''
Phys. Rev. D \textbf{85}, 054503 (2012)
%doi:10.1103/PhysRevD.85.054503
[arXiv:1111.1710 [hep-lat]].

\bibitem{Aoki:2006we}
Y.~Aoki, G.~Endrodi, \textit{et al}, 
%``The Order of the quantum chromodynamics transition predicted by the standard model of particle physics,''
Nature \textbf{443}, 675-678 (2006)
%doi:10.1038/nature05120
[arXiv:hep-lat/0611014 [hep-lat]].



\bibitem{Heinz:2013th}
U.~Heinz and R.~Snellings,
%``Collective flow and viscosity in relativistic heavy-ion collisions,''
Ann. Rev. Nucl. Part. Sci. \textbf{63}, 123-151 (2013)
%doi:10.1146/annurev-nucl-102212-170540
[arXiv:1301.2826 [nucl-th]].

\bibitem{Kolb:2003dz}
P.~F.~Kolb and U.~W.~Heinz,
%``Hydrodynamic description of ultrarelativistic heavy ion collisions,''
[arXiv:nucl-th/0305084 [nucl-th]].

\bibitem{Song:2010mg}
H.~Song, S.~A.~Bass, U.~Heinz, T.~Hirano and C.~Shen,
%``200 A GeV Au+Au collisions serve a nearly perfect quark-gluon liquid,''
Phys. Rev. Lett. \textbf{106}, 192301 (2011)
[erratum: Phys. Rev. Lett. \textbf{109}, 139904 (2012)]
%doi:10.1103/PhysRevLett.106.192301
[arXiv:1011.2783 [nucl-th]].


\bibitem{Gale:2012rq}
C.~Gale, S.~Jeon, B.~Schenke, P.~Tribedy and R.~Venugopalan,
%``Event-by-event anisotropic flow in heavy-ion collisions from combined Yang-Mills and viscous fluid dynamics,''
Phys. Rev. Lett. \textbf{110}, no.1, 012302 (2013)
%doi:10.1103/PhysRevLett.110.012302
[arXiv:1209.6330 [nucl-th]].

\bibitem{STAR:2004jwm}
J.~Adams \textit{et al.} [STAR],
%``Azimuthal anisotropy in Au+Au collisions at s(NN)**(1/2) = 200-GeV,''
Phys. Rev. C \textbf{72}, 014904 (2005)
%doi:10.1103/PhysRevC.72.014904
[arXiv:nucl-ex/0409033 [nucl-ex]].

\bibitem{ALICE:2011ab}
K.~Aamodt \textit{et al.} [ALICE],
%``Higher harmonic anisotropic flow measurements of charged particles in Pb-Pb collisions at $\sqrt{s_{NN}}$=2.76 TeV,''
Phys. Rev. Lett. \textbf{107}, 032301 (2011)
%doi:10.1103/PhysRevLett.107.032301
[arXiv:1105.3865 [nucl-ex]].


\bibitem{Matsui:1986dk}
T.~Matsui and H.~Satz,
%``$J/\psi$ Suppression by Quark-Gluon Plasma Formation,''
Phys. Lett. B \textbf{178}, 416-422 (1986)
%doi:10.1016/0370-2693(86)91404-8


\bibitem{Braun-Munzinger:2000csl}
P.~Braun-Munzinger and J.~Stachel,
%``(Non)thermal aspects of charmonium production and a new look at J / psi suppression,''
Phys. Lett. B \textbf{490}, 196-202 (2000)
%doi:10.1016/S0370-2693(00)00991-6
[arXiv:nucl-th/0007059 [nucl-th]].

\bibitem{Zhu:2004nw}
X.~l.~Zhu, P.~f.~Zhuang and N.~Xu,
%``J/psi transport in QGP and p(t) distribution at SPS and RHIC,''
Phys. Lett. B \textbf{607}, 107-114 (2005)
%doi:10.1016/j.physletb.2004.12.023
[arXiv:nucl-th/0411093 [nucl-th]].

\bibitem{Yan:2006ve}
L.~Yan, P.~Zhuang and N.~Xu,
%``Competition between J / psi suppression and regeneration in quark-gluon plasma,''
Phys. Rev. Lett. \textbf{97}, 232301 (2006)
%doi:10.1103/PhysRevLett.97.232301
[arXiv:nucl-th/0608010 [nucl-th]].

\bibitem{Liu:2010ej}
Y.~Liu, B.~Chen, N.~Xu and P.~Zhuang,
%``$\Upsilon$ Production as a Probe for Early State Dynamics in High Energy Nuclear Collisions at RHIC,''
Phys. Lett. B \textbf{697}, 32-36 (2011)
%doi:10.1016/j.physletb.2011.01.026
[arXiv:1009.2585 [nucl-th]].

\bibitem{Grandchamp:2003uw}
L.~Grandchamp, R.~Rapp and G.~E.~Brown,
%``In medium effects on charmonium production in heavy ion collisions,''
Phys. Rev. Lett. \textbf{92}, 212301 (2004)
%doi:10.1103/PhysRevLett.92.212301
[arXiv:hep-ph/0306077 [hep-ph]].

\bibitem{Zhao:2011cv}
X.~Zhao and R.~Rapp,
%``Medium Modifications and Production of Charmonia at LHC,''
Nucl. Phys. A \textbf{859}, 114-125 (2011)
%doi:10.1016/j.nuclphysa.2011.05.001
[arXiv:1102.2194 [hep-ph]].

\bibitem{Blaizot:2015hya}
J.~P.~Blaizot, D.~De Boni, P.~Faccioli and G.~Garberoglio,
%``Heavy quark bound states in a quark\textendash{}gluon plasma: Dissociation and recombination,''
Nucl. Phys. A \textbf{946}, 49-88 (2016)
%doi:10.1016/j.nuclphysa.2015.10.011
[arXiv:1503.03857 [nucl-th]].


\bibitem{Krouppa:2015yoa}
B.~Krouppa, R.~Ryblewski and M.~Strickland,
%``Bottomonia suppression in 2.76 TeV Pb-Pb collisions,''
Phys. Rev. C \textbf{92}, no.6, 061901 (2015)
%doi:10.1103/PhysRevC.92.061901
[arXiv:1507.03951 [hep-ph]].

\bibitem{Yao:2020xzw}
X.~Yao, W.~Ke, Y.~Xu, S.~A.~Bass and B.~M\"uller,
%``Coupled Boltzmann Transport Equations of Heavy Quarks and Quarkonia in Quark-Gluon Plasma,''
JHEP \textbf{01}, 046 (2021)
%doi:10.1007/JHEP01(2021)046
[arXiv:2004.06746 [hep-ph]].

\bibitem{Yao:2020eqy}
X.~Yao and T.~Mehen,
%``Quarkonium Semiclassical Transport in Quark-Gluon Plasma: Factorization and Quantum Correction,''
JHEP \textbf{02}, 062 (2021)
%doi:10.1007/JHEP02(2021)062
[arXiv:2009.02408 [hep-ph]].

\bibitem{Wen:2022utn}
L.~Wen, X.~Du, S.~Shi and B.~Chen,
%``Probe the color screening in proton-nucleus collisions with complex potentials,''
Chin. Phys. C \textbf{46}, 114102 (2022)
%doi:10.1088/1674-1137/ac7fe6
[arXiv:2205.07520 [nucl-th]].

\bibitem{Shi:2017qep}
W.~Shi, W.~Zha and B.~Chen,
%``Charmonium Coherent Photoproduction and Hadroproduction with Effects of Quark Gluon Plasma,''
Phys. Lett. B \textbf{777}, 399-405 (2018)
%doi:10.1016/j.physletb.2017.12.055
[arXiv:1710.00332 [nucl-th]].

\bibitem{NA50:2000mfb}
M.~C.~Abreu \textit{et al.} [NA50],
%``Transverse momentum distributions of J / psi, psi-prime, Drell-Yan and continuum dimuons produced in Pb Pb interactions at the SPS,''
Phys. Lett. B \textbf{499}, 85-96 (2001)
%doi:10.1016/S0370-2693(01)00019-3

\bibitem{PHENIX:2006gsi}
A.~Adare \textit{et al.} [PHENIX],
%``$J/\psi$ Production vs Centrality, Transverse Momentum, and Rapidity in Au+Au Collisions at $\sqrt{s_{NN}} = 200$ GeV,''
Phys. Rev. Lett. \textbf{98}, 232301 (2007)
%doi:10.1103/PhysRevLett.98.232301
[arXiv:nucl-ex/0611020 [nucl-ex]].



\bibitem{ALICE:2015jrl}
J.~Adam \textit{et al.} [ALICE],
%``Differential studies of inclusive J/\ensuremath{\psi} and \ensuremath{\psi}(2S) production at forward rapidity in Pb-Pb collisions at $ \sqrt{s_{\mathrm{NN}}}=2.76 $ TeV,''
JHEP \textbf{05}, 179 (2016)
%doi:10.1007/JHEP05(2016)179
[arXiv:1506.08804 [nucl-ex]].

\bibitem{ALICE:2016flj}
J.~Adam \textit{et al.} [ALICE],
%``J/$\psi$ suppression at forward rapidity in Pb-Pb collisions at $\mathbf{\sqrt{s_{{\rm NN}}} = 5.02}$ TeV,''
Phys. Lett. B \textbf{766}, 212-224 (2017)
%doi:10.1016/j.physletb.2016.12.064
[arXiv:1606.08197 [nucl-ex]].
\bibitem{Thews:2000rj}
R.~L.~Thews, M.~Schroedter and J.~Rafelski,
%``Enhanced $J/\psi$ production in deconfined quark matter,''
Phys. Rev. C \textbf{63}, 054905 (2001)
%doi:10.1103/PhysRevC.63.054905
[arXiv:hep-ph/0007323 [hep-ph]].

\bibitem{Greco:2003vf}
V.~Greco, C.~M.~Ko and R.~Rapp,
%``Quark coalescence for charmed mesons in ultrarelativistic heavy ion collisions,''
Phys. Lett. B \textbf{595}, 202-208 (2004)
%doi:10.1016/j.physletb.2004.06.064
[arXiv:nucl-th/0312100 [nucl-th]].

\bibitem{Andronic:2003zv}
A.~Andronic, P.~Braun-Munzinger, K.~Redlich and J.~Stachel,
%``Statistical hadronization of charm in heavy ion collisions at SPS, RHIC and LHC,''
Phys. Lett. B \textbf{571}, 36-44 (2003)
%doi:10.1016/j.physletb.2003.07.066
[arXiv:nucl-th/0303036 [nucl-th]].

\bibitem{Fries:2008hs}
R.~J.~Fries, V.~Greco and P.~Sorensen,
%``Coalescence Models For Hadron Formation From Quark Gluon Plasma,''
Ann. Rev. Nucl. Part. Sci. \textbf{58}, 177-205 (2008)
%doi:10.1146/annurev.nucl.58.110707.171134
[arXiv:0807.4939 [nucl-th]].

\bibitem{Du:2015wha}
X.~Du and R.~Rapp,
%``Sequential Regeneration of Charmonia in Heavy-Ion Collisions,''
Nucl. Phys. A \textbf{943}, 147-158 (2015)
%doi:10.1016/j.nuclphysa.2015.09.006
[arXiv:1504.00670 [hep-ph]].

\bibitem{Chen:2017duy}
B.~Chen and J.~Zhao,
%``Bottomonium Continuous Production from Unequilibrium Bottom Quarks in Ultrarelativistic Heavy Ion Collisions,''
Phys. Lett. B \textbf{772}, 819-824 (2017)
%doi:10.1016/j.physletb.2017.07.054
[arXiv:1704.05622 [nucl-th]].

\bibitem{Zhao:2017yan}
J.~Zhao and B.~Chen,
%``Strong diffusion effect of charm quarks on J /$\psi$ production in Pb\textendash{}Pb collisions at the LHC,''
Phys. Lett. B \textbf{776}, 17-21 (2018)
%doi:10.1016/j.physletb.2017.11.014
[arXiv:1705.04558 [nucl-th]].



\bibitem{CMS:2012bms}
S.~Chatrchyan \textit{et al.} [CMS],
%``Suppression of non-prompt $J/\psi$, prompt $J/\psi$, and Y(1S) in PbPb collisions at $\sqrt{s_{NN}}=2.76$ TeV,''
JHEP \textbf{05}, 063 (2012)
%doi:10.1007/JHEP05(2012)063
[arXiv:1201.5069 [nucl-ex]].

\bibitem{ALICE:2013olq}
B.~Abelev \textit{et al.} [ALICE],
%``D meson elliptic flow in non-central Pb-Pb collisions at $\sqrt{s_{\rm NN}}$ = 2.76TeV,''
Phys. Rev. Lett. \textbf{111}, 102301 (2013)
%doi:10.1103/PhysRevLett.111.102301
[arXiv:1305.2707 [nucl-ex]].


\bibitem{Cao:2013ita}
S.~Cao, G.~Y.~Qin and S.~A.~Bass,
%``Heavy-quark dynamics and hadronization in ultrarelativistic heavy-ion collisions: Collisional versus radiative energy loss,''
Phys. Rev. C \textbf{88}, 044907 (2013)
%doi:10.1103/PhysRevC.88.044907
[arXiv:1308.0617 [nucl-th]].

\bibitem{Cao:2018ews}
S.~Cao, G.~Coci, S.~K.~Das, W.~Ke, S.~Y.~F.~Liu, S.~Plumari, T.~Song, Y.~Xu, J.~Aichelin and S.~Bass, \textit{et al.}
%``Toward the determination of heavy-quark transport coefficients in quark-gluon plasma,''
Phys. Rev. C \textbf{99}, no.5, 054907 (2019)
%doi:10.1103/PhysRevC.99.054907
[arXiv:1809.07894 [nucl-th]].

\bibitem{He:2012df}
M.~He, R.~J.~Fries and R.~Rapp,
%``$\mathbf{D_s}$-Meson as Quantitative Probe of Diffusion and Hadronization in Nuclear Collisions,''
Phys. Rev. Lett. \textbf{110}, no.11, 112301 (2013)
%doi:10.1103/PhysRevLett.110.112301
[arXiv:1204.4442 [nucl-th]].

\bibitem{He:2019vgs}
M.~He and R.~Rapp,
%``Hadronization and Charm-Hadron Ratios in Heavy-Ion Collisions,''
Phys. Rev. Lett. \textbf{124}, no.4, 042301 (2020)
%doi:10.1103/PhysRevLett.124.042301
[arXiv:1905.09216 [nucl-th]].



\bibitem{Chen:2016mhl}
B.~Chen,
%``Elliptic flow as a probe for $\psi(2S)$ production mechanism in relativistic heavy ion collisions,''
Phys. Rev. C \textbf{95}, no.3, 034908 (2017)
%doi:10.1103/PhysRevC.95.034908
[arXiv:1608.02173 [nucl-th]].

\bibitem{Liu:2009nb}
Y.~p.~Liu, Z.~Qu, N.~Xu and P.~f.~Zhuang,
%``J/psi Transverse Momentum Distribution in High Energy Nuclear Collisions at RHIC,''
Phys. Lett. B \textbf{678}, 72-76 (2009)
%doi:10.1016/j.physletb.2009.06.006
[arXiv:0901.2757 [nucl-th]].




\bibitem{Chen:2021akx}
B.~Chen, L.~Jiang, X.~H.~Liu, Y.~Liu and J.~Zhao,
%``$X(3872)$ Production in Relativistic Heavy-Ion Collisions,''
Phys. Rev. C \textbf{105}, 054901 (2022)
%doi:10.1103/PhysRevC.105.054901
[arXiv:2107.00969 [hep-ph]].

\bibitem{Cao:2015hia}
S.~Cao, G.~Y.~Qin and S.~A.~Bass,
%``Energy loss, hadronization and hadronic interactions of heavy flavors in relativistic heavy-ion collisions,''
Phys. Rev. C \textbf{92}, no.2, 024907 (2015)
%doi:10.1103/PhysRevC.92.024907
[arXiv:1505.01413 [nucl-th]].


\bibitem{Rapp:2018qla}
R.~Rapp, P.~B.~Gossiaux, A.~Andronic, R.~Averbeck, S.~Masciocchi, A.~Beraudo, E.~Bratkovskaya, P.~Braun-Munzinger, S.~Cao and A.~Dainese, \textit{et al.}
%``Extraction of Heavy-Flavor Transport Coefficients in QCD Matter,''
Nucl. Phys. A \textbf{979}, 21-86 (2018)
%doi:10.1016/j.nuclphysa.2018.09.002
[arXiv:1803.03824 [nucl-th]].

\bibitem{Zhao:2020jqu}
J.~Zhao, K.~Zhou, S.~Chen and P.~Zhuang,
%``Heavy flavors under extreme conditions in high energy nuclear collisions,''
Prog. Part. Nucl. Phys. \textbf{114}, 103801 (2020)
%doi:10.1016/j.ppnp.2020.103801
[arXiv:2005.08277 [nucl-th]].


\bibitem{Guo:2000nz}
X.~f.~Guo and X.~N.~Wang,
%``Multiple scattering, parton energy loss and modified fragmentation functions in deeply inelastic e A scattering,''
Phys. Rev. Lett. \textbf{85}, 3591-3594 (2000)
%doi:10.1103/PhysRevLett.85.3591
[arXiv:hep-ph/0005044 [hep-ph]].

\bibitem{Zhang:2003wk}
B.~W.~Zhang, E.~Wang and X.~N.~Wang,
%``Heavy quark energy loss in nuclear medium,''
Phys. Rev. Lett. \textbf{93}, 072301 (2004)
%doi:10.1103/PhysRevLett.93.072301
[arXiv:nucl-th/0309040 [nucl-th]].

\bibitem{Cacciari:2001td}
M.~Cacciari, S.~Frixione and P.~Nason,
%``The p(T) spectrum in heavy flavor photoproduction,''
JHEP \textbf{03}, 006 (2001)
%doi:10.1088/1126-6708/2001/03/006
[arXiv:hep-ph/0102134 [hep-ph]].

\bibitem{Huovinen:2009yb}
P.~Huovinen and P.~Petreczky,
%``QCD Equation of State and Hadron Resonance Gas,''
Nucl. Phys. A \textbf{837}, 26-53 (2010)
%doi:10.1016/j.nuclphysa.2010.02.015
[arXiv:0912.2541 [hep-ph]].

\bibitem{Schenke:2010nt}
B.~Schenke, S.~Jeon and C.~Gale,
%``(3+1)D hydrodynamic simulation of relativistic heavy-ion collisions,''
Phys. Rev. C \textbf{82}, 014903 (2010)
%doi:10.1103/PhysRevC.82.014903
[arXiv:1004.1408 [hep-ph]].


\bibitem{Chen:2015iga}
B.~Chen,
%``Detailed rapidity dependence of $J/\psi$ production at energies available at the Large Hadron Collider,''
Phys. Rev. C \textbf{93}, no.5, 054905 (2016)
%doi:10.1103/PhysRevC.93.054905
[arXiv:1510.07466 [hep-ph]].

\bibitem{Chen:2013wmr}
B.~Chen, Y.~Liu, K.~Zhou and P.~Zhuang,
%``$\psi^\prime$ Production and $B$ Decay in Heavy Ion Collisions at {LHC},''
Phys. Lett. B \textbf{726}, 725-728 (2013)
%doi:10.1016/j.physletb.2013.09.036
[arXiv:1306.5032 [nucl-th]].





\bibitem{Peskin:1979va}
M.~E.~Peskin,
%``Short Distance Analysis for Heavy Quark Systems. 1. Diagrammatics,''
Nucl. Phys. B \textbf{156}, 365-390 (1979)
%doi:10.1016/0550-3213(79)90199-8

\bibitem{Bhanot:1979vb}
G.~Bhanot and M.~E.~Peskin,
%``Short Distance Analysis for Heavy Quark Systems. 2. Applications,''
Nucl. Phys. B \textbf{156}, 391-416 (1979)
%doi:10.1016/0550-3213(79)90200-1



\bibitem{ALICE:2012vgf}
B.~Abelev \textit{et al.} [ALICE],
%``Anisotropic flow of charged hadrons, pions and (anti-)protons measured at high transverse momentum in Pb-Pb collisions at $\sqrt{s_{NN}}$=2.76 TeV,''
Phys. Lett. B \textbf{719}, 18-28 (2013)
%doi:10.1016/j.physletb.2012.12.066
[arXiv:1205.5761 [nucl-ex]].


\bibitem{ALICE:2012vup}
B.~Abelev \textit{et al.} [ALICE],
%``Inclusive $J/\psi$ production in $pp$ collisions at $\sqrt{s} = 2.76$ TeV,''
Phys. Lett. B \textbf{718}, 295-306 (2012)
[erratum: Phys. Lett. B \textbf{748}, 472-473 (2015)]
%doi:10.1016/j.physletb.2012.10.078
[arXiv:1203.3641 [hep-ex]].

\bibitem{ALICE:2012vpz}
B.~Abelev \textit{et al.} [ALICE],
%``Measurement of prompt $J/\psi$ and beauty hadron production cross sections at mid-rapidity in $pp$ collisions at $\sqrt{s} = 7$ TeV,''
JHEP \textbf{11}, 065 (2012)
%doi:10.1007/JHEP11(2012)065
[arXiv:1205.5880 [hep-ex]].

\bibitem{Chen:2018kfo}
B.~Chen,
%``Thermal production of charmonia in Pb-Pb collisions at $\sqrt {s_{NN}} = $ 5.02 TeV,''
Chin. Phys. C \textbf{43}, no.12, 124101 (2019)
%doi:10.1088/1674-1137/43/12/124101
[arXiv:1811.11393 [nucl-th]].

\bibitem{Chen:2016dke}
B.~Chen, T.~Guo, Y.~Liu and P.~Zhuang,
%``Cold and Hot Nuclear Matter Effects on Charmonium Production in p+Pb Collisions at LHC Energy,''
Phys. Lett. B \textbf{765}, 323-327 (2017)
%doi:10.1016/j.physletb.2016.12.021
[arXiv:1607.07927 [nucl-th]].




\bibitem{Eskola:2009uj}
K.~J.~Eskola, H.~Paukkunen and C.~A.~Salgado,
%``EPS09: A New Generation of NLO and LO Nuclear Parton Distribution Functions,''
JHEP \textbf{04}, 065 (2009)
%doi:10.1088/1126-6708/2009/04/065
[arXiv:0902.4154 [hep-ph]].

\bibitem{He:2021zej}
M.~He, B.~Wu and R.~Rapp,
%``Collectivity of J/\ensuremath{\psi} Mesons in Heavy-Ion Collisions,''
Phys. Rev. Lett. \textbf{128}, no.16, 162301 (2022)
%doi:10.1103/PhysRevLett.128.162301
[arXiv:2111.13528 [nucl-th]].




\bibitem{Alver:2010gr}
B.~Alver and G.~Roland,
%``Collision geometry fluctuations and triangular flow in heavy-ion collisions,''
Phys. Rev. C \textbf{81}, 054905 (2010)
[erratum: Phys. Rev. C \textbf{82}, 039903 (2010)]
%doi:10.1103/PhysRevC.82.039903
[arXiv:1003.0194 [nucl-th]].

\bibitem{Miller:2003kd}
M.~Miller and R.~Snellings,
%``Eccentricity fluctuations and its possible effect on elliptic flow measurements,''
[arXiv:nucl-ex/0312008 [nucl-ex]].

\bibitem{Miller:2007ri}
M.~L.~Miller, K.~Reygers, S.~J.~Sanders and P.~Steinberg,
%``Glauber modeling in high energy nuclear collisions,''
Ann. Rev. Nucl. Part. Sci. \textbf{57}, 205-243 (2007)
%doi:10.1146/annurev.nucl.57.090506.123020
[arXiv:nucl-ex/0701025 [nucl-ex]].

\bibitem{Kharzeev:2001gp}
D.~Kharzeev and E.~Levin,
%``Manifestations of high density QCD in the first RHIC data,''
Phys. Lett. B \textbf{523}, 79-87 (2001)
%doi:10.1016/S0370-2693(01)01309-0
[arXiv:nucl-th/0108006 [nucl-th]].

\bibitem{Alver:2010dn}
B.~H.~Alver, C.~Gombeaud, M.~Luzum and J.~Y.~Ollitrault,
%``Triangular flow in hydrodynamics and transport theory,''
Phys. Rev. C \textbf{82}, 034913 (2010)
%doi:10.1103/PhysRevC.82.034913
[arXiv:1007.5469 [nucl-th]].

\bibitem{Qiu:2011iv}
Z.~Qiu and U.~W.~Heinz,
%``Event-by-event shape and flow fluctuations of relativistic heavy-ion collision fireballs,''
Phys. Rev. C \textbf{84}, 024911 (2011)
%doi:10.1103/PhysRevC.84.024911
[arXiv:1104.0650 [nucl-th]].

\bibitem{Zhao:2021voa}
J.~Zhao, B.~Chen and P.~Zhuang,
%``Charmonium triangular flow in high energy nuclear collisions,''
Phys. Rev. C \textbf{105}, no.3, 034902 (2022)
%doi:10.1103/PhysRevC.105.034902
[arXiv:2112.00293 [hep-ph]].

\bibitem{STAR:2008jgm}
B.~I.~Abelev \textit{et al.} [STAR],
%``System-size independence of directed flow at the Relativistic Heavy-Ion Collider,''
Phys. Rev. Lett. \textbf{101}, 252301 (2008)
%doi:10.1103/PhysRevLett.101.252301
[arXiv:0807.1518 [nucl-ex]].

\bibitem{STAR:2014clz}
L.~Adamczyk \textit{et al.} [STAR],
%``Beam-Energy Dependence of the Directed Flow of Protons, Antiprotons, and Pions in Au+Au Collisions,''
Phys. Rev. Lett. \textbf{112}, no.16, 162301 (2014)
%doi:10.1103/PhysRevLett.112.162301
[arXiv:1401.3043 [nucl-ex]].

\bibitem{Chen:2019qzx}
B.~Chen, M.~Hu, H.~Zhang and J.~Zhao,
%``Probe the tilted Quark-Gluon Plasma with charmonium directed flow,''
Phys. Lett. B \textbf{802}, 135271 (2020)
%doi:10.1016/j.physletb.2020.135271
[arXiv:1910.08275 [nucl-th]].

\bibitem{STAR:2019clv}
J.~Adam \textit{et al.} [STAR],
%``First Observation of the Directed Flow of $D^{0}$ and $\overline{D^0}$ in Au+Au Collisions at $\sqrt{s_{\rm NN}}$ = 200 GeV,''
Phys. Rev. Lett. \textbf{123}, no.16, 162301 (2019)
%doi:10.1103/PhysRevLett.123.162301
[arXiv:1905.02052 [nucl-ex]].






%%TTTT



\end{thebibliography}
\end{document}